\documentclass[%
reprint,
superscriptaddress,
showpacs,preprintnumbers,
showkeys,longbibliography,
nofootinbib,
amsmath,amssymb,
aps,
pra,
]{revtex4-1}
\usepackage{graphicx}
\usepackage{dcolumn}
\usepackage{bm}
\usepackage{hyperref}
\usepackage{amsfonts}
\hypersetup{%
 colorlinks = true,
 breaklinks = true,
 citecolor = blue,
 linkcolor = blue,
 urlcolor = blue,
 linkbordercolor = 0 0 0,
 pdfborder= 0 0 0,
}

\usepackage[pagewise, mathlines]{lineno}
\usepackage{physics}
\usepackage{xcolor}

\usepackage{multirow}
\usepackage{mwe}
\usepackage{adjustbox}
\usepackage{lipsum}
\usepackage{booktabs}
\usepackage{soul}
\usepackage{bm}
\usepackage[toc,page]{appendix}
\usepackage{comment}

\begin{document}

\title{Surface trap with adjustable ion couplings for scalable and parallel gates}
 
 \author{Y. Suleimen}
 \email{suleimenelnur@gmail.com}
 \affiliation{Russian Quantum Center, Russia, Skolkovo, Moscow region, 121205}

 \author{A. Podlesnyy}
 \email{a.podlesnyy@rqc.ru}
 \affiliation{Russian Quantum Center, Russia, Skolkovo, Moscow region, 121205}

\author{L. A. Akopyan}
 \affiliation{Russian Quantum Center, Russia, Skolkovo, Moscow region, 121205}

 \author{N. Sterligov}
 \affiliation{Russian Quantum Center, Russia, Skolkovo, Moscow region, 121205}

\author{O. Lakhmanskaya}
 \affiliation{Russian Quantum Center, Russia, Skolkovo, Moscow region, 121205}

\author{E. Anikin}
 \affiliation{Russian Quantum Center, Russia, Skolkovo, Moscow region, 121205}

\author{A. Matveev}
 \affiliation{Russian Quantum Center, Russia, Skolkovo, Moscow region, 121205}
 
\author{K. Lakhmanskiy}
 \affiliation{Russian Quantum Center, Russia, Skolkovo, Moscow region, 121205}


\begin{abstract}
We describe the design and operation of a surface-electrode Paul trap for parallel entangling gate implementation. In particular, we demonstrate the possibility of separating or coupling ion motion by adjusting the DC-voltages on a set of electrodes and show the possibility of parallel MS-gate operations for specific voltage configurations. We verify the scalability of this approach and characterize the performance of these gates in the presence of the finite phonon mode occupation and of the finite drift of the phonon frequencies. Additionally, we investigate how the number of ions per individual trapping site and anharmonic potential terms affect the coupling between the wells.

\end{abstract}
\maketitle



\section{Introduction} 

Trapped ions remain one of the leading technology platforms for large-scale quantum computers (QCs) \cite{Cirac2000, Zhu2022, Cheng2023}. In particular, few-ion-qubit systems have successfully been demonstrated with high performance \cite{Harty2014, Ballance2016, Clark2021, Pogoverolov2021, Wright2019,Baldwin2020,Benhelm2008,Choi2014,Gaebler2016,Hahn2019,Harty2016,Hughes2020,Leung2018,Milne2020,Schafer2018,Srinivas2021,Tan2013,Wang2020,Weidt2016,Zarantonello2019}. 

Practical realizations of QCs require the ability to increase the number of simultaneously trapped ions while maintaining the ability to control and measure them individually with high fidelity \cite{Shor1994, Ambainis2003, Santha2008,Harrow2009, Paudel2022, Lu2012}. Single linear arrays of ions encounter significant limitations in that respect. This is primarily due to the increase in the number of motional modes and their density with the length of the crystal. As a result speed and performance of two-qubit gates between ions in a chain generally decrease as the total number of ions grows \cite{Bruzewicz2019,Pogoverolov2021, Wright2019}. 

A promising approach to these issues is to break a single long ion chain into segments \cite{kielpinski2002architecture, Cirac2000} arranged in a surface trap. Each such segment or module can trap a restricted number of ions to maintain high fidelity and high-speed operations.
In such a modular approach, each subsystem can be built and tested independently, has a particular and well-defined functionality, and is compatible with the other subsystems. Therefore, surface traps have the potential to offer a fast and high-fidelity method to distribute quantum information over a many-ion array \cite{mehta2020integrated,Craig2021,srinivas2021high}.

The challenge, then, becomes how to move quantum information between the modules. The entanglement between the ions in separated segments can be achieved in different ways: through ion transport \cite{kielpinski2002architecture}, through effective spin-spin interactions \cite{Cirac2000}, through shared electrically floating electrodes \cite{Daniilidis_2009, An2022}, controlled orientation of the secular modes \cite{Hakelberg2019}, or through photonic interconnects \cite{Moehring2007, Monroe2014, Krutyanskiy2019, Schupp2021}. 
Another method demonstrated for two trapping regions relies on the tuning of the Coulomb interaction of  ions in two separate wells \cite{Harlander2011, brown2011coupled}. Followed by the proposed method new entanglement schemes for qubit spins were suggested and tested \cite{ wilson2014tunable}. Here we assume standard M$\o$lmer-S$\o$rensen (MS) entangling gate \cite{Molmer1999} and describe a trap design with multiple individual trapping sites utilizing the proposed method of the motional coupling \cite{Harlander2011, brown2011coupled} to manipulate the structure of the phonon mode spectrum. We characterize this new surface trap design 
and show that it is capable of coupling the ion motion in the selected sites by adjusting the DC-voltages on a set of electrodes. Moreover, sufficient control over the ion motion in our trap offers an alternative method to generate programmable interactions and connectivity of the ion qubits beyond the power law without tweezers \cite{Espinoza2021}. New design also helps to get control over MS-gate performance, since the phonon mode structure and their spectral density may have a profound influence on the fidelity of the MS gate \cite{Olsacher2020, Sosnova2021}. 

The paper is organized as follows. At first, we demonstrate and characterize a variety of phonon mode spectra for different voltage sets for two types of ions: calcium and beryllium. Then, we focus on the capability to unite the ions into segments with unique phonon mode frequencies to perform parallel MS-gate operations and verify the scalability of this approach.  We also model the fidelity of the MS-gate operation for a segment comprising two ions accounting for the two main contributions to the gate infidelity: the limited frequency difference between phonon frequencies of the segments and the finite drift of the phonon frequencies expected in the experiment. We account for the possible issues related to anomalous heating rate typical for such traps \cite{Turchette2000, Deslauriers2006, Hite2017, Boldin2017} and determine the optimum range of voltages and frequencies to negotiate this effect. At last, we investigate how the number of ions per individual trapping site and anharmonic potential terms affect the coupling between the two wells.

\begin{figure}[b]
\includegraphics[width=3.375in]{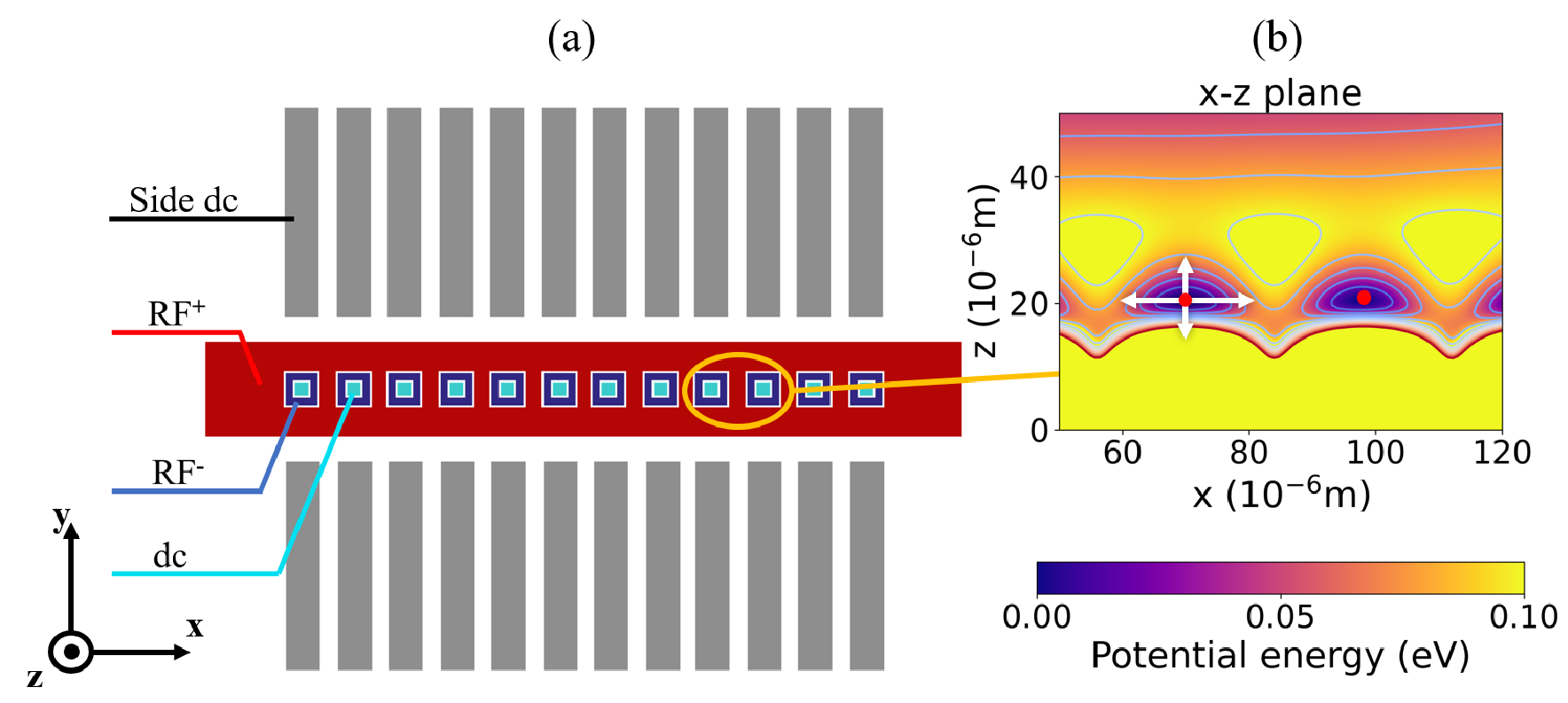}
\caption[font = footnotesize]{\label{fig:Trapwithfield}(a) The proposed surface ion trap geometry forming 12 potential wells. The red RF electrode (labelled RF$^+$) and 12 dark-blue RF electrodes (labelled RF$^-$) with $\pi$ phase delay form ion confinement with predetermined height. Twelve light-blue dc electrodes (labelled dc) in the middle of squares are utilized for secular frequency optimization. The side dc electrodes (labelled Side dc), depicted in gray, are essential for stray field compensation.
(b) Potential distribution above two individual traps demonstrating a trap depth of 60-100\,meV. The principal axes of oscillation are illustrated by white arrows. Pseudopotential is shown for Ca ions in the configuration with grounded dc electrodes and V$_{RF}$=80\,V, $\Omega_{RF}$=2$\pi\times$110\,MHz. The potential for Be ions is similar for the respective RF drive.}
\end{figure}

\section{Trap structure}

In this section, we present a trap design (see Fig. \ref{fig:Trapwithfield}) allowing manipulation of the normal mode spectrum with an electrode voltage set. 
From the practical point of view, in planar traps, it is more convenient to change dc voltages than RF drive. Therefore, we suggest the structure combining individual and linear surface trap designs schematically shown in Fig. \ref{fig:Trapwithfield}(a). It includes a single RF$^+$ electrode with 12 square notches aligned along the $X$ axis (axial direction), and creating separate potential wells to confine the ions. Each notch constitutes the RF$^-$ electrode and the central optimization dc electrode. The phase delay $\pi$ between RF$^+$ and RF$^-$ electrodes helps to increase the ion height above the surface. The ions are contained only in ten central trapping sites. The first and the last notches in the array produce too shallow potential wells and are used as endcaps. 
On both sides of the RF$^+$ electrode, there are 12 side dc electrodes (see Fig. \ref{fig:Trapwithfield}(a)). These electrodes are necessary for stray field and micromotion compensation \cite{Narayanan2011} and, additionally, can be used for the fine-tuning and optimization of the secular frequencies. In general, this design can be scaled to an arbitrary number of trapping sites. First, we focus on and characterize the design with 12 notches, and in Section \ref{sec:spectral} we discuss the scalability of our approach.

To choose the design we account for several technical constraints of the system including trap depth $U_{depth}$, and for a possibility to entangle ions via MS gate (see Sec. \ref{sec:Coupling strength and connectivity}). MS gate can be only implemented for ions which are motionally coupled. Thus, for trap design optimization it is necessary to quantify the motional coupling strength between two interacting ions of mass $m$ in two separate individual wells $i$ and $j$. We use frequency splitting between normal modes, because it has a close relation to the Coulomb coupling term introduced in \cite{brown2011coupled, wilson2014tunable} (see details in appendix \ref{appendix:Coupling Hamiltonian}):

\begin{equation}
\label{eq:interaction}
\Omega^{ij}_{I}=\frac{e^2}{2\pi\epsilon_0 m\sqrt{\omega_{i}\omega_{j}}d_{ion}^3},
\end{equation}
where $\epsilon_0$ is the vacuum electric permittivity, $\omega_{i,j}$ are uncoupled secular frequencies, respectively. 

Namely, when the uncoupled secular frequencies of two wells are the same, the Coulomb coupling strength can be expressed through the frequency difference between the center-of-mass (COM) and breathing normal modes of a pair of ions:
\begin{equation}
\label{eq:interaction_fr}
\Omega^{ij}_I = 2\pi \Delta f^{ij},
\end{equation}
Although Eq.\eqref{eq:interaction} and Eq.\eqref{eq:interaction_fr} are obtained for only two trapping sites \cite{brown2011coupled, wilson2014tunable}, we show numerically  in Appendix \ref{appendix:Coupling Hamiltonian}, that $\Delta f^{ij}$ can be also used to characterize the coupling of the two selected wells in a multi trapping zone architecture. 

For the fixed $U_{depth}$, the secular frequency and the coupling strength between the ions are $\omega_{sec} \propto d_{ion}^{-1}$ and $\Omega_I \propto m^{-1}d_{ion}^{-2}$, respectively.
Therefore, to preserve trap depth and at the same time to increase the coupling strength between the ions in two different trapping sites, the distances between trapping regions have to be sufficiently small. In particular, the widths of the dc and RF$^-$ electrodes are 6 {\textmu}m, the width of the RF$^+$ electrode between two notches is 10 {\textmu}m, and the width of the gaps between them is 2 {\textmu}m.

Fig. \ref{fig:Trapwithfield}(b) demonstrates the simulated pseudopotential above the two individual traps. To achieve the global trap depth of about $100$ meV we optimize the geometry of the surface trap and RF parameters. For one $^{40}$Ca$^+$ ($^{9}$Be$^+$) ion per trapping site, it is achieved at $V_{RF}$ = 80(85)\,V (from 0 to Peak) and $\Omega_{RF}$ = 2$\pi\times$110(240)\,MHz for both RF$^{\pm}$. The ion trapping occurs at the height of 20 {\textmu}m in each site. We estimate the numerical aperture for the optical access of about 0.55 at this height, which implies the necessity of the addressing based on the integrated photonics \cite{Ivory2021}. 
Concerning connectivity, the distance of $d_{ion} = 28$ {\textmu}m between the individual potential minima has been demonstrated to be sufficient to share motional quanta \cite{brown2011coupled}. 

\begin{figure}
\includegraphics[width=3.375in]{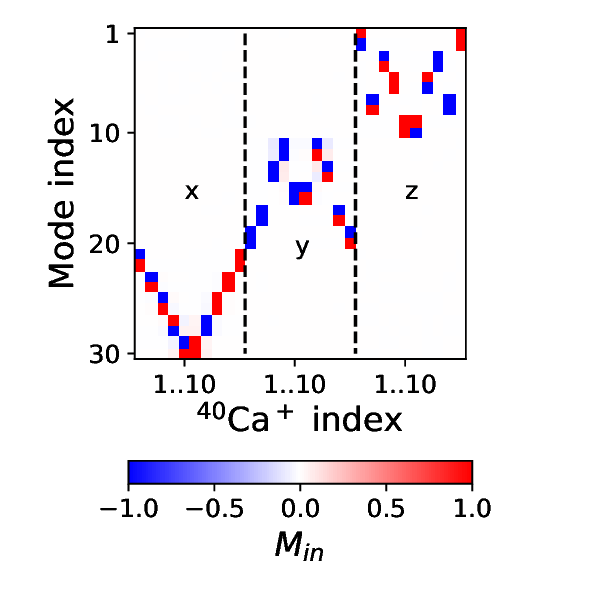}

\begin{ruledtabular}
\begin{tabular}{lcccccc}
Mode index ($n$) & 1 & 10 & 11 & 20 & 21 & 30 \\
\colrule
$f_n$ (MHz) & 28.5 & 28.3 & 17.9 & 17.5 & 11.0 & 10.4 \\
\end{tabular}
\end{ruledtabular}

\caption{\label{fig:Caw28dc03N} The normal mode interaction matrix of the ion crystal formed in the trap in Fig. \ref{fig:Trapwithfield} with grounded dc electrodes. The matrix is divided into three parts, representing modes, corresponding to the respective principal axis of oscillation. The cell's color represents the normalized interaction strength $M_{in}$ from Eq. \eqref{eq:intermatrix} between ion $i$ and normal mode $n$. For $z$-axis, the difference in secular frequency ranges from 0.02 to 0.14 MHz. On the table the frequencies $f_n$ of the respective normal modes $n$ are presented. 
 }
\end{figure} 

\begin{figure*}
\begin{tabular}{llll}
\includegraphics[width=2.3in]{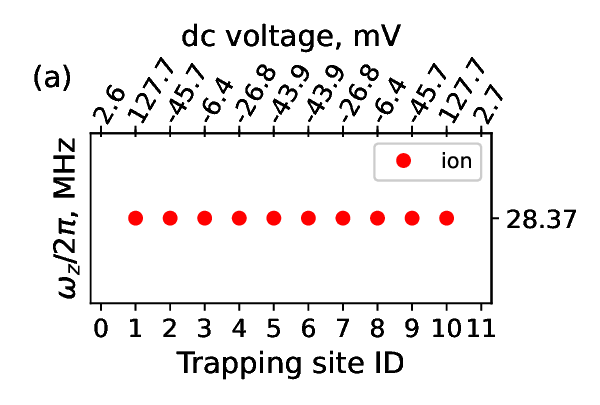}&
\includegraphics[width=2.3in]{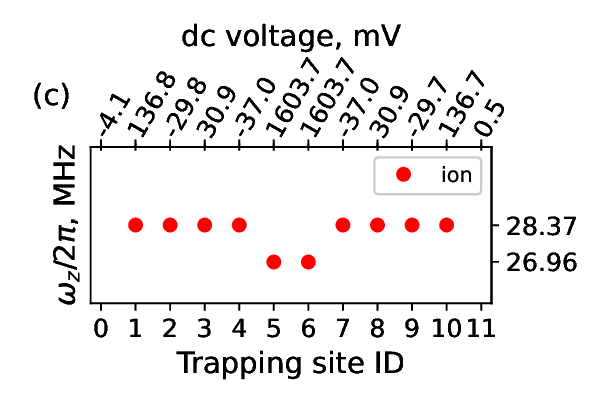}&
\includegraphics[width=2.3in]{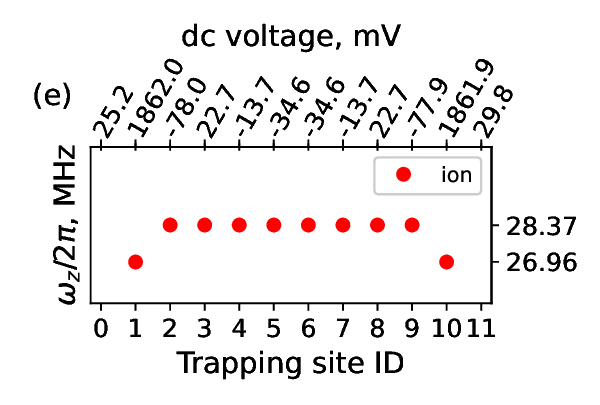}\\
    \addlinespace
\includegraphics[width=1.99in]{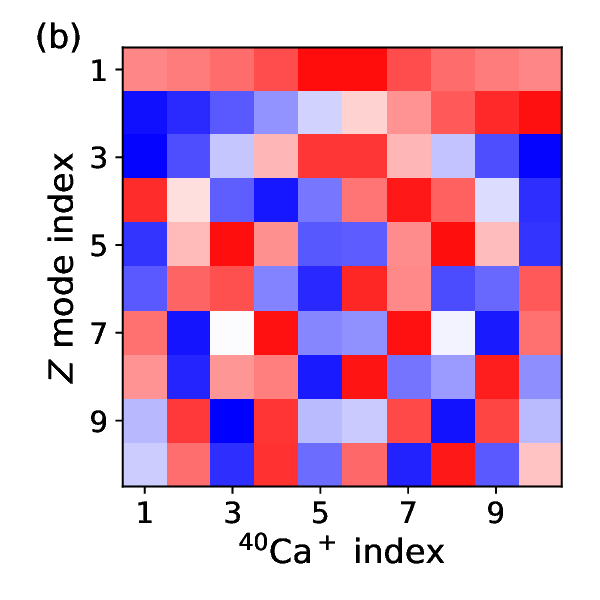}&
\includegraphics[width=1.99in]{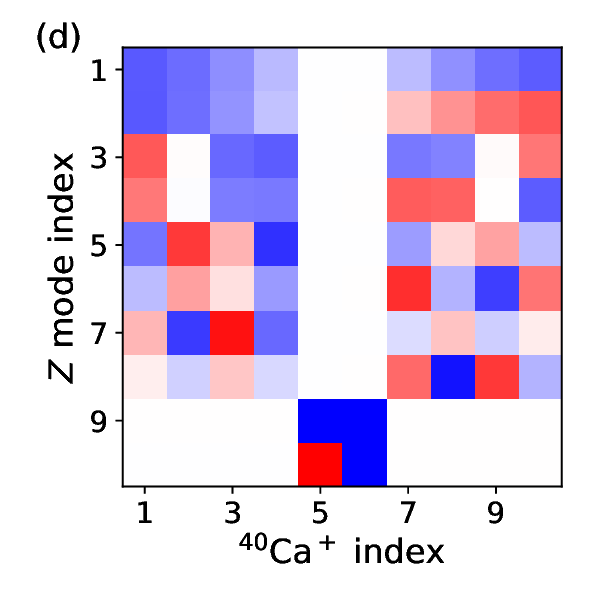}&
\includegraphics[width=1.99in]{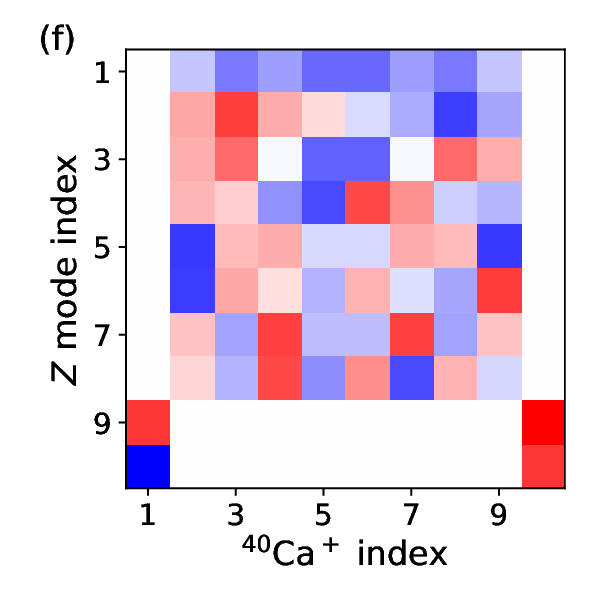}\\
\end{tabular}
\includegraphics[width=5in]{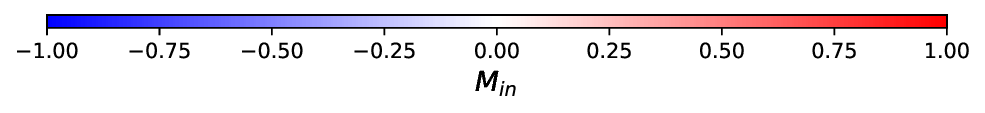}

\begin{ruledtabular}
\begin{tabular}{lccccccccc}
Mode index ($n$) & 2 & 3 & 4 & 5 & 6 & 7 & 8 & 9 & 10 \\
\colrule
(b) $f_{n-1} - f_n$ (Hz) & 27 & 21 & 52 & 39 & 42 & 44 & 22 & 38 & 6  \\
(d) $f_{n-1} - f_n$ (Hz) & 2 & 83 & 3 & 95 & 2 & 70 & 1 & $1.41\times 10^6$ & $\boldsymbol{141}$  \\
(f) $f_{n-1} - f_n$ (Hz) & 33 & 39 & 96 & 61 & 6 & 51 & 13 & $1.41\times 10^6$ & $\boldsymbol{0.2}$  \\
\end{tabular}
\end{ruledtabular}

\caption{\label{fig:Caw28cenpin} Spectra for ten $^{40}$Ca$^+$ ions sitting in the trap shown in Fig. \ref{fig:Trapwithfield} for different dc voltage configurations. Traps with indices 0 and 11 do not contain ions and are used as end caps to decrease the required voltages on the inner trapping sites. Panels (a, c, e) demonstrate radial secular frequencies and the required dc voltage configuration. The respective panels (b, d, f) show interaction matrices $M$ with the colors corresponding to the normalized strength of the ion interaction calculated according to Eq. \eqref{eq:intermatrix}. The table presents the differences between two neighboring normal mode frequencies $f_{n-1}-f_n$.
}
\end{figure*}

The choice of the RF drive frequencies $\Omega_{RF}$ and voltages has a strong impact on several parameters. The potential barrier between two neighboring traps in pseudopotential approximation is $U_{depth} \propto \omega^2_xd_{ion}^2$. Therefore, high secular frequencies $\omega_{x}\propto V_{RF}/\Omega_{RF}$ are required to preserve sufficient trap depth during miniaturization. On the other hand, stability parameters for the surface ion traps scale proportionally to $\omega_{x}$:  $q\propto \omega_{x}/\Omega_{RF}$ and must not exceed $0.908$ to support stable motion \cite{Shaikh2012}. Consequently, the trapping potential depth scales as $U_{depth}\propto V_{RF}^2/\Omega_{RF}^2$. Technical limitations restrict the magnitude of $V_{RF}$ and, in general, lead to a shallow depth in chip-based ion traps. The optimal $\Omega_{RF}$ for preserving stable ion motion and sufficiently low $V_{RF}$ are derived taking into account these features. 

\begin{figure}[b]
\begin{tabular}{c}
\includegraphics[width=3in]{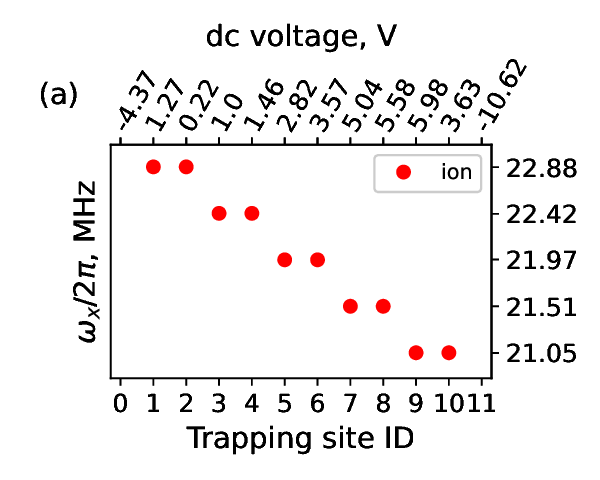} \\
\includegraphics[width=3in]{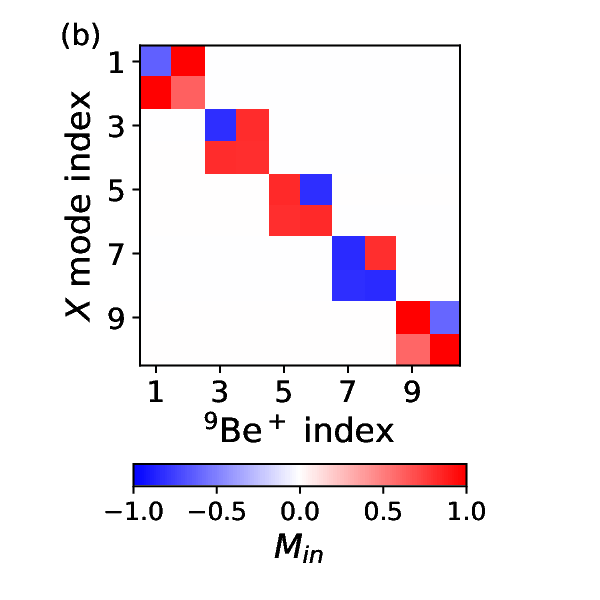}
\end{tabular}

\begin{ruledtabular}
\begin{tabular}{lccccc}
Mode index ($n$) & 1 & 3 & 5 & 7 & 9  \\
\colrule
$f_{n}-f_{n+1}$ (Hz) & 1860 & 1622 & 1654 & 1649 & 1763 \\
\end{tabular}
\end{ruledtabular}

\caption{\label{fig:Bew28axialladder}
Spectral segmentation optimized for parallel entangling gate implementation for ten $^9$Be$^+$ ions sitting in the trap shown in Fig.  \ref{fig:Trapwithfield}. Traps with indices 0 and 11 do not contain ions and are used as end caps to decrease the required voltages on the inner trapping sites. Panel (a) shows axial secular frequencies and the required dc voltage configuration. Panel (b) shows the respective interaction matrix $M$ with the colors corresponding to the normalized strength of the ion-to-mode interaction calculated according to Eq. \eqref{eq:intermatrix}. 
}
\end{figure}

The magnitude of typical parasitic fields in surface ion traps ranges around 20--30 Vm$^{-1}$. Under these conditions, the electrical barrier of 60--70 meV for the ion separation should be resistant to dual to solitary trapping site rearrangement \cite{Pholz200}. The potential well depth decreases with higher voltages on the dc electrodes. We identified that the range of operating voltages between 0 and 6 V on the dc central electrodes in the trapping sites keeps the trap depth $U_{depth}$ above 50 meV, which is above the energy of the molecules in the residual gas (at room temperature $U_{bcg}\propto 40$ meV). The latter minimizes the ion loss due to the collisions with background gas molecules \cite{SouersFeb1977}.
At ultra-high vacuum, we estimate the trapped ion's lifespan as $\tau=\frac{kT_{300K}ln(2)}{P\sigma}\sqrt{\frac{\pi m_{H_2}}{8kT_{300K}}}\propto36$ $min$, which is sufficient to execute quantum processes. 

To analyze secular phonon mode spectra and to find secular frequencies for different voltage sets we compute pseudo-potential Hessian $A$ \cite{Home2013} with the following components: 
\begin{equation}
\label{eq:hessian}
    (A_{ij})_k = \delta_{ij}m_i(\omega^2_i/2)_k + (C_{ij})_k,
\end{equation}
where the first term represents the hessian for the trapping potential with secular frequency $\omega_i$ and $C_{ij}$ is a Coulomb interaction hessian, $\delta_{ij}$ -- Kronecker delta, $m_i$ -- the mass of the ion \emph{i}, and $k$ denotes the principal axis of oscillations ($x,y, z$). The normal mode vectors $Q_n(t)$ and frequencies of the ion oscillations in the Coulomb crystal are computed by diagonalization of matrix $A$ numerically. This implies the transformation from the original deviations from equilibrium positions $q_i(t)$ to $Q_n(t)$ using transformation matrix elements $b_{in}$:
\begin{equation}
\label{eq:normalmode}
    Q_n(t) = \sum^{N}_{n=1}b_{in}q_i(t).
\end{equation}
To analyze the coupling strength between an ions $i$ and a phonon mode $n$ we use normalized mode matrix $M$ with the following elements:
\begin{equation}
\label{eq:intermatrix}
    M_{in}=b_{in}/\text{max}(b_{in}),
\end{equation}
Further, we refer to this matrix M as an interaction matrix.
The secular frequencies of each trapping site are computed by diagonalizing the Hessian of the corresponding confining potential $\Psi$ as  $\omega_{i}^2 = em_i^{-1}\emph{Eig}\left( {\partial^2\Psi}/{\partial r_n\partial r_m} \right)$,  with $e$ being the elementary charge \cite{Shmied2009}.

For the grounded dc electrodes we obtained the following secular frequencies in the central potential well: $(\omega_{x},\omega_{y},\omega_{z})_{Be^+} = \left(22.60, 38.74, 61.33\right) \times 2\pi$ MHz for Be ions and $(\omega_{x},\omega_{y},\omega_{z})_{Ca^+} = \left(10.44, 17.9, 28.34\right) \times 2\pi$ MHz for Ca ions.

\section{Motional couplings and scalability} 
\label{sec:spectral}

Here we demonstrate the possibility of forming an arbitrary configuration of ion couplings through the appropriate choice of the voltages on the central dc electrodes and the scalability of this approach.
For this, we model ion motion and calculate the phonon mode spectra using the molecular dynamics Python package described in Appendix \ref{appendix:simulation}. The details on the simulations are also presented in Appendix \ref{appendix:simulation}. 

For all the results discussed in the paper, we observe small variations of the ion heights: $\delta z_i\leq 1.5\times 10^{-2} z_{mean}$. Nevertheless, we do not observe the coupling between the normal modes corresponding to different principal axes, which is demonstrated in Fig. \ref{fig:Caw28dc03N}.
Therefore, further, we demonstrate the results for only one of the axes. All calculations of normal modes, however, were performed without the linear chain approximation. 

When the dc electrodes are all grounded, the symmetric structure of the RF$^{\pm}$ and the dc electrodes forms the symmetric set of secular frequencies obeying the relation: $\omega_{i}\ =\omega_{10-i}$ for $i$ ranging from 1 to 5. As a result, the pairs of the ions with indices $(i, 10-i)$, $i=1..5$ are motionally coupled as demonstrated in Fig. \ref{fig:Caw28dc03N}. Each pair shares the same axial and radial phonon modes, thus forming a spectral segment. Further, we use this term to refer to the restricted set of normal modes, attributed to the strongly coupled ions, which motion can be considered independently from the rest of the ions in the crystal. In Fig. \ref{fig:Caw28dc03N} the phonon mode frequency differences between the spectral segments range between 0.02 and 0.14 MHz. The interaction between the uncoupled ions is absent because of the significant secular frequency difference $|\Delta^{i,j}|\gg \Omega_{I}^{ij}$ (see Appendix \ref{appendix:Coupling Hamiltonian}), where $\Delta^{i,j}=\omega_{i} - \omega_{j}$. The resonant condition is then expressed as $\Delta^{i,j}\ll \Omega_{I}^{ij}$, with a resonance peak being at $\Delta^{i,j} = 0$.
 
Fig. \ref{fig:Caw28cenpin} shows some of the phonon mode configurations achieved at specific sets of voltages leading to the formation of spectral segments. To achieve those configurations we developed a numerical optimization procedure for the voltage sets (described in Appendix \ref{appendix:DCoptimization}). Figs. \ref{fig:Caw28cenpin}(a-b) demonstrate all-to-all connectivity between the ions through radial \emph{z} modes, whereas Figs. \ref{fig:Caw28cenpin}(c-f) show the possibility of uniting arbitrary ions into a segment. Namely, the coupling between the ions and the modes of the central segment in Fig. \ref{fig:Caw28cenpin}(d) is significantly larger than the coupling of other ions to these modes:
\begin{equation}\label{Eq:couplings}
    \begin{aligned}
        \frac{b_{i_{rest}, n_{pin}}}{b_{i_{pin},n_{pin}}}\leq10^{-4},
    \end{aligned}
\end{equation}
where $b$ are the elements of the transformation matrix determined in Eq. \eqref{eq:normalmode}. The index $i$ corresponds to the ion, and $n$ -- to the normal mode, while index notation $pin$ denotes the ions/modes, which are united in a considered spectral segment, and $rest$ denotes the rest. This relation proves that the considered ion pair forms a spectral segment. To form such a segment the ions shall sit in the trapping sites with the same $\omega_{sec}$. 

\begin{figure*}
\begin{tabular}{ccc}
(a) & (b) & (c)\\
\includegraphics[width=2.in]{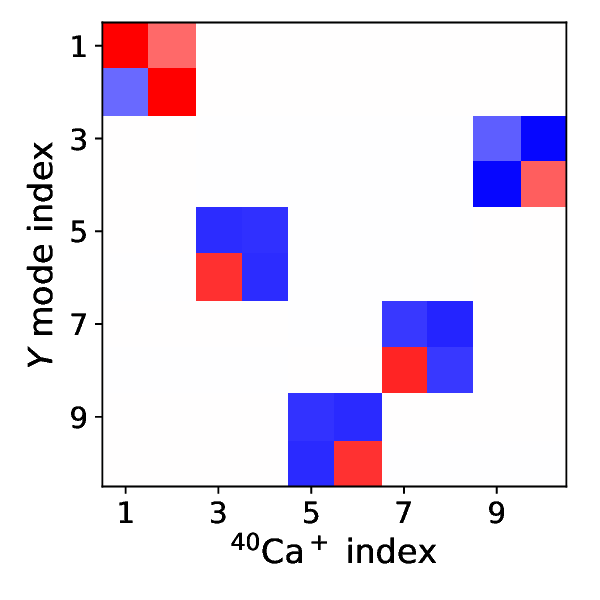}&
\includegraphics[width=2.in]{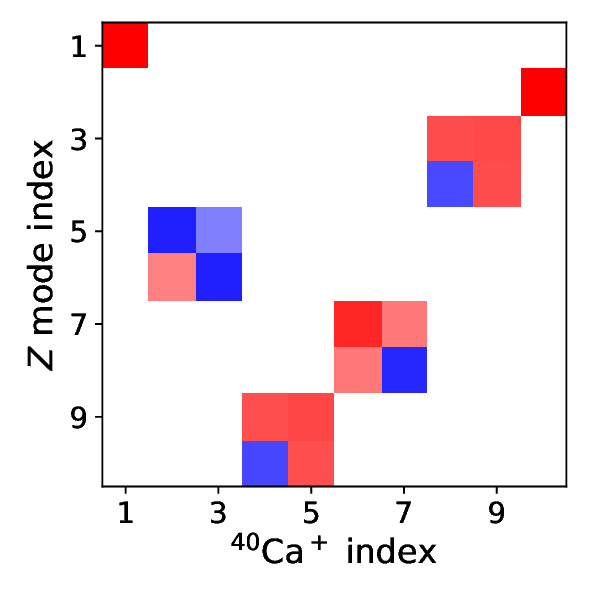}&
\includegraphics[width=2.in]{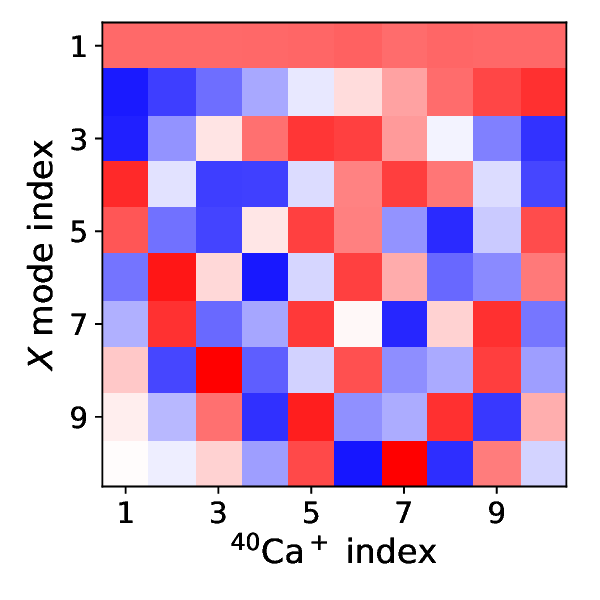}
\end{tabular}
\includegraphics[width=5in]{colorbar.eps}
\caption{\label{fig:heisenberg} Interaction matrices of ten $^{40}$Ca$^+$ ions obtained with the use of gray and blue dc electrodes shown in Fig. \ref{fig:Trapwithfield}. The colors correspond to the normalized strength of the ion-to-mode interaction calculated according to Eq. \eqref{eq:intermatrix}. Panels (a) and (b) demonstrate segmentation of the phonon mode structure in radial directions while panel (c) shows the axial interaction matrix with all-to-all connectivity for the same ion chain and trap configuration.}
\end{figure*}

\begin{figure}[b]
\includegraphics[width=3.375in]{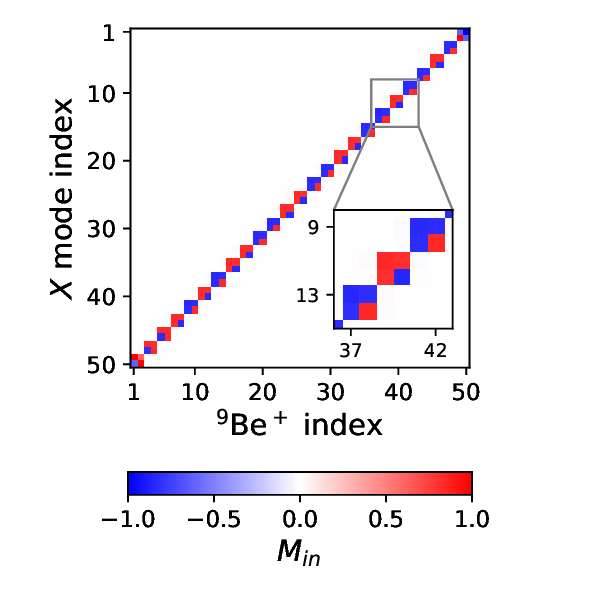}

\begin{ruledtabular}
\begin{tabular}{lcccc}
Mode index ($n$) & $2n-1$ & $2n$ & $2n+1 $ \\
\colrule
$f_{n}$ (kHz) & $f_{2n-1}$ & $f_{2n-1} + 1.6$ & $f_{2n-1} + 44.7 $\\
\end{tabular}
\end{ruledtabular}

\caption{\label{fig:Bew2850ladder} Interaction matrix of normal modes of a 50-ion crystal, in the scaled ion trap from Fig.  \ref{fig:Trapwithfield}. The optimal configuration of voltages allows the forming of 25 independent ion pairs suitable for parallel gates. The table depicts the general rule for the normal mode frequency in this configuration.}
\end{figure}

In Figs. \ref{fig:Caw28cenpin}(e-f), the central segment unites all the central ions but the two on the sides (with indices 1 and 10). The interaction strength between the ions within those segments is significantly reduced compared to the previous case: $\Omega_I^{1,10} = 2\pi\times 1 $ Hz for the side ions. This is because the coupling strength decreases with the inter ion distance (see Eq. \eqref{eq:interaction}). In particular, the coupling strength in a segment, comprising two ions, sitting in non-neighboring trapping sites, separated by another one is approximately 8 times smaller compared to the case with neighboring ions. 
Therefore, the best performance of the two-qubit entangling operations can be achieved with the segments consisting of the couple of ions in neighboring trapping depicted in Figs. \ref{fig:Caw28cenpin}(c-d) and \ref{fig:Bew28axialladder}. 

For the segments comprising more than two ions in Fig. \ref{fig:Caw28cenpin}, $M_{i,COM}$-components for the COM mode are not equal. Similar feature is observed in mixed-species ion chains \cite{Sosnova2021}. The limited accuracy of the optimization procedure leads to the mismatch of the secular frequencies of the order of 20 Hz within trapping sites united in a spectral segment. As a result, the resonance condition is not satisfied: $\Delta^{i,j}\lesssim \Omega_{I}^{ij}$. On the other side it is hard to achieve a control over secular frequencies better than 20 Hz too. In fact to meet the resonant condition one needs a dc voltage supply with ppm accuracy.
To address this issue we simplify the resonant condition increasing the coupling strength.
Namely, the separation of the normal mode frequencies scales as $f_n-f_{COM} \propto \omega_{sec}^{-1}m^{-1}$ and, therefore, can be significantly improved for lighter ions like beryllium and for axial modes as $\omega_{ax}<\omega_{rad}$. Additionally, axial modes have higher separation even for equal secular frequencies, due to the difference in axial and radial hessians of the potential. 
Further we consider parallel operations with axial normal modes, exhibiting greatest coupling strength.

Fig. \ref{fig:Bew28axialladder} illustrates a possible segmentation of the ion phonon spectra in the axial direction. This configuration of the spectra is particularly interesting for parallel gate operations and will be characterized and discussed in detail in section \ref{sec:parallel}. Namely, ten ions form five segments with five distinct secular frequencies $\omega_x$, whereas in each segment good coupling is achieved. The highest frequency separation was achieved for the segmented Be ion pair in Fig. \ref{fig:Bew28axialladder}. For ions $i=1,2$ the axial mode frequency separation is $\Delta f= 1.8$ kHz.

The side dc electrodes (Fig. \ref{fig:Trapwithfield}) can be also used to modify the phonon mode spectrum of an ion crystal. This allows optimizing secular frequencies of motion at all orthogonal principal oscillation axes in parallel. 
In particular, in Fig. \ref{fig:heisenberg}  
the interaction between neighboring ions is achieved by respective segmentation of the radial ($y$ and $z$) modes on pairs.
Each segment is characterized  by two normal mode frequencies: COM and stretch. In the axial direction $x$ all-to-all connectivity is retained.

To scale the trap one needs to add notches. This implies the increase of the number of electrodes: the number of dc electrodes scales as $3N$, and the number of RF$^-$ electrodes scales as $N$. The amount of side dc electrodes can be reduced, since they are used only for stray field compensation. To achieve reliable control on secular frequencies the control of dc voltages with ppm accuracy is necessary. We verify the capability to create segmented spectra  in the trap with 50 notches (for Be ions). We added the necessary number of notches N and elongated the central RF$^-$ electrode by $N\times d_{ion} = N\times 28$ {\textmu}m to keep the axial secular frequency unchanged. Fig. \ref{fig:Bew2850ladder} demonstrates the result for the optimized voltage set forming 25 segments uniting the nearest neighbors. The required voltage set lies in the operating regime. The secular frequency  difference between the segments is $\Delta^{2k, 2k+1}/2\pi = 44.7$ kHz, while the average frequency difference of the phonon modes within a segment is $\Delta f = 1.6$ kHz. For this case 
$b_{i_{rest}, n_{pin}}/b_{i_{pin},n_{pin}}\leq0.02$ (see Eq. \eqref{Eq:couplings}). Segmentation can be further improved by increasing $\Delta^{2k, 2k+1}$, which is limited by the breakdown voltages in the present design.
Several effects have to be accounted for while implementing this procedure. The increase in the upper voltage bound diminishes the individual well depth. The lower voltage bound can be only extended when the stability region of the trap is expanded, which can be achieved by a simultaneous increase in $\Omega_{RF}$ and $V_{RF}$ to preserve the trap depth.

\section{Motional coupling strength and entanglement}
\label{sec:Coupling strength and connectivity}

\begin{table*}
\caption{\label{tab:Rabivalues} Minimum MS-gate duration $t_g$ required to achieve theoretical unit fidelity on the selected ion pair. We assume MS-gate with constant amplitude and with Rabi frequency of 2 MHz and minimize $t_g$ by adjusting laser detuning $\delta$. The coupled ions are specified by their trapping site IDs and the normal mode configuration, referenced by the corresponding figure. The coupling strength is characterized by frequency difference of the COM mode $f_{COM}$ and the closest breathing mode of the selected segment $f_{br}$, according to Eq. \eqref{eq:interaction_fr}.}
\begin{ruledtabular}
\begin{tabular}{cccccccccc}
Ion           & Site ID & Configuration     & $|f_{COM}-f_{br}|$ (Hz) & $t_g$ ($\mu$s) & Mode axis & $f_{COM}$ (MHz) & $\delta$ (Hz) \\
\colrule \\
$^{40}$Ca$^+$ & 5-6     & Segmented pair (Fig. \ref{fig:Caw28cenpin}(d))   & 142                  & 224            & $z$       & 26.98          & -6500     \\  
              & 2-3     & Parallel gates (Fig. \ref{fig:heisenberg}(b))   & 159                     & 229            & $z$       & 27.81          & -6365      \\  
\colrule  \\
$^{9}$Be$^+$  & 5-6     & Segmented pair (Fig. \ref{fig:symmasimm}(d))   & 1534                 & 20             & $x$       & 21.57          & 739   \\  
              & 3-4    & Parallel gates (Fig. \ref{fig:Bew28axialladder})   & 1622                    & 24            & $x$       & 22.42          & 3737 \\  
\end{tabular}
\end{ruledtabular}
\end{table*}

In this section we verify the feasibility to perform MS gates on the proposed trap. For this we evaluate MS gate duration $t_g$ required for the ideal theoretical gate. To illustrate its relation to  the mode splitting $\Delta f$ and, therefore, to the motional coupling strength, we first derive the generalized relation between $t_g$ and $\Delta f$ considering only two trapping sites. We assume, that the difference between COM and breathing modes is small $2\pi\Delta f = |\omega_{COM} - \omega_{br}| \ll \delta$, where $\delta = \mu - \omega_{COM}$ is laser detuning from COM-mode. In this case the conventional MS gate scheme without segmentation is applicable. For such a scheme, the Rabi frequency $\Omega$ is constant and the set of Eq. \eqref{eq: alpha and chi} reduces to the following: 
\begin{equation}
\label{eq:quantised_ham}
\begin{aligned}
& \delta t_g = 2\pi K, \quad K \in \mathbf{Z},\\
& 2\eta^2\Omega^2 2\pi\Delta f \left | -\frac{t_g}{4\delta^2} \left ( 1+\cos{\delta t_g} \right ) + \frac{2\sin{\delta t_g}}{\delta^3}  \right | = \frac{\pi}{4}.\\
\end{aligned}
\end{equation}
Finally, the relation between gate duration and mode splitting reads:
\begin{equation}\label{eq:gate_params}
\frac{\eta^2\Omega^2}{\delta^2}  \Delta f \, t_g =  \frac{1}{8}.
\end{equation}
According to this formula, either the Rabi frequency, or the duration of the entangling operation should increase with the increased distance between the traps. Eq. \eqref{eq:gate_params} also explains how gate parameters depend on the ion mass. Namely,  the optimal gate time decreases with decreasing mass of the ion, as the frequency difference $\Delta f$ and the square of Lamb-Dicke parameter are inversely proportional to the mass of the ion.

In general case the gate parameters can be found numerically using the segmentation approach \cite{Choi2014, Zhu2006EPL} described in appendix \ref{appendix:MSevolution}. We examine mode structures presented in figures \ref{fig:Caw28cenpin}(d), \ref{fig:heisenberg}(b), \ref{fig:Bew28axialladder}, \ref{fig:symmasimm}(c) and assume a MS laser pulse with 5 intervals. The maximum Rabi frequency on an interval was restricted by an experimentally feasible value of 2\,MHz \cite{Hendricks2008, Lo2014, Burd2023}. The choice of MS gate parameters is not unique, in Table \ref{tab:Rabivalues} we present those allowing to minimize gate duration and to satisfy Lamb-Dicke approximation. Gate duration below 1 ms can be considered as reasonably good based on the coherence times of Be and Ca qubits (exceed 1\,s \cite{Langer2005, Bruzewicz2019, Bermudez2017}). Additionally, within this time heating rate and secular frequency drifts have a moderate effect on the gate performance (see Sec. \ref{sec:Errors}). 
 
We verified that harmonic potential approximation holds for the cases with frequency splitting above 100 Hz. As a result, to correctly estimate duration of MS gate for the more distant trapping sites one has to account for the anharmonicity of the trapping potential which is beyond the scope of this publication. Overall, we do not exclude the possibility to couple more distant trapping sites especially for light atoms like Be and for larger laser intensities ($\Omega$).

\section{Noise sources}\label{sec:Errors}
The dominant sources of gate errors in surface ion traps are the drift of normal modes and anomalous heating rate \cite{mehta2020integrated}. The drift is typically in the order of several kHz per minute \cite{mehta2020integrated} and may lead to incomplete decoupling of qubit spins with normal modes at the end of the entangling gate. 

The anomalous heating increases the amplitude of the motion of the ions and the motional-state decoherence \cite{Turchette2000, Deslauriers2006, Hite2017, Boldin2017}. The anomalous heating strongly depends on multiple parameters such as spatial separation between the ions and the electrode surfaces, ion mass, secular frequencies \cite{Sedlacek2018, Lakhmanskiy2019}, temperature, and spectral density of the electric field.
Based on the results in \cite{Sedlacek2018} and \cite{Ivory2021}, we assume the axial heating rate in our cases to be below 100-s phonons per second at cryogenic temperatures and below 1000 phonons per second at room temperature. For a typical gate time of about 200 {\textmu}s, this implies less than 0.2 phonons per gate. This is a fair assumption, even though the heating of the COM mode increases as the number of trapped ions rises. Namely, performing gates with the stretch mode allows decreasing the heating rate by a factor of hundred: $\dot{n}_{str}/\dot{n}_{com}\propto 0.01$  \cite{Maciej2021}. Therefore, in the upcoming sections we assume that the noise does not change the amount of phonons during the gate. Instead, we estimate the effect of the finite phonon populations on the two qubit-gate infidelity. 

\section{Robustness of parallel gates}
\label{sec:parallel}

In this section, we focus on the main beneficial feature of the phonon mode spectrum configuration depicted in Fig. \ref{fig:Bew28axialladder}: the possibility to perform parallel two-qubit entangling gates. Namely, we compute the performance of the entangling Mølmer–Sørensen (MS) operation \cite{Molmer1999} between the ions within one of the segments and characterize its robustness against the noise types described in section \ref{sec:Errors}. 
In this section we demonstrate the results only for Be ions, though similar quantitative behavior of fidelities and Rabi frequency is achieved also with Ca ions.

\begin{figure}[b]
\includegraphics[width=3.375in]{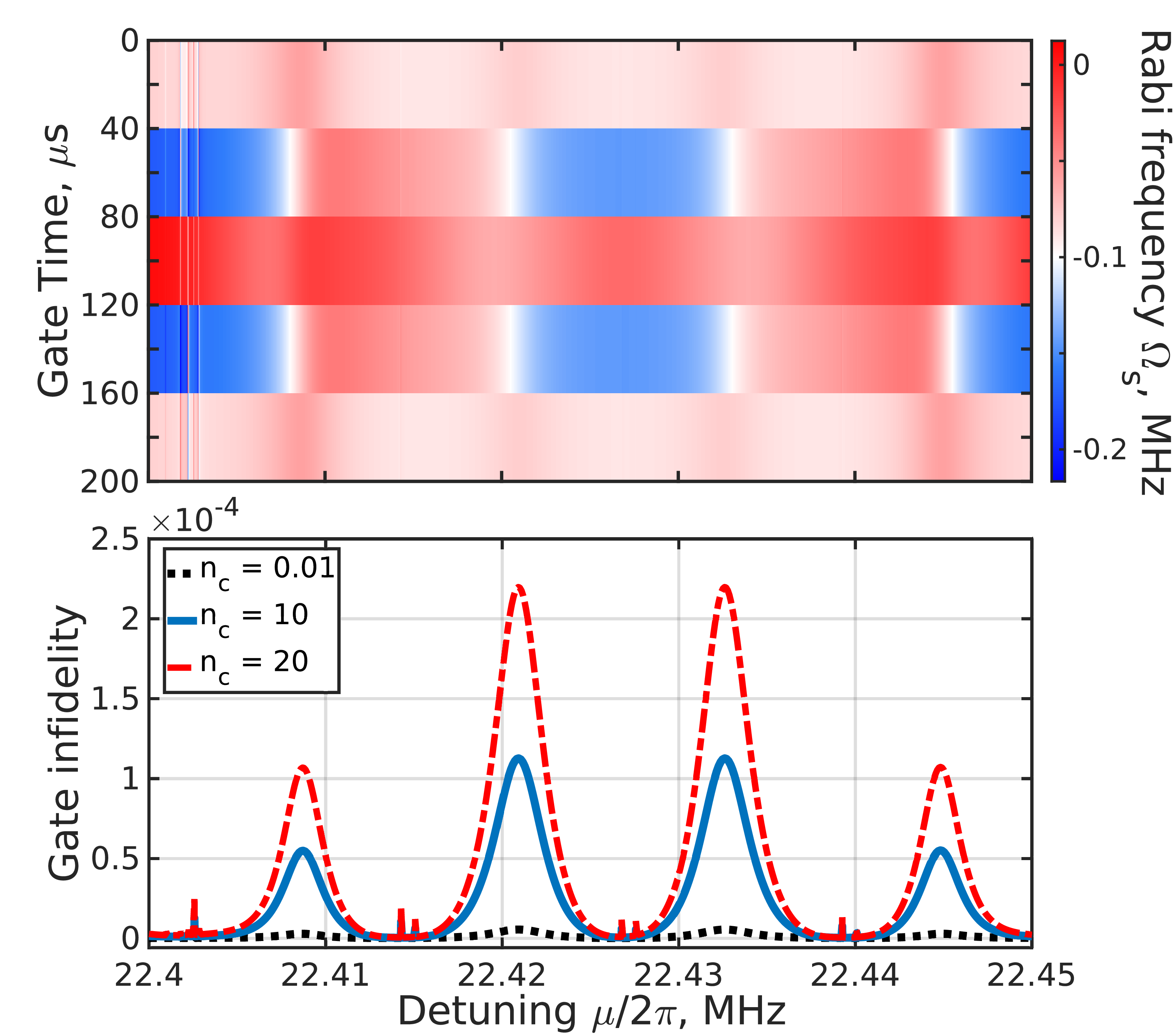}
\caption{\label{fig:RFreqsHeatmap}Parameters for entangling gate on the united pair of ions (with indices 3 and 4) for trap configuration in Fig. \ref{fig:Bew28axialladder}. Panel (a) shows in-time segmented Rabi frequency $\Omega_s$ in MHz for two neighboring $^9$Be$^+$ ions. (b) MS-gate infidelity over detuning $\mu$ for two neighboring $^9$Be$^+$ ions for different initial average normal mode occupations. The initial average occupation for the ion pair center-of-mass mode is specified in the legend. The drift rates for all the modes are taken as 1 MHz/min. The narrow local infidelity maxima correspond to the normal frequencies of ions 3 and 4 and half of their sum, all three are repeated with frequency $5/2t_g$. Rabi frequency instabilities coincide with gate infidelity local maxima.}
\end{figure}

First of all we account for and mitigate gate errors due to the presence of many collective modes of motion and their closeness (see Appendix \ref{appendix:MSevolution}) with the aid of optimum modulation of the laser amplitude as suggested in \cite{Choi2014, Zhu2006, Sosnova2021, Zhu2006EPL}.  The optimal parameters of the gate for two Be ions $i = 3$ and 4 are shown in Fig. \ref{fig:RFreqsHeatmap}(a). In general, the ion chain comprising $N$ ions requires to split the gate pulse onto $2N+1$ intervals with specific Rabi frequencies $\Omega_s$ on each interval. But the spectral configuration in Fig. \ref{fig:Bew28axialladder} allows reducing the number of intervals to five per ion segment, which significantly simplifies the optimization routine. Namely, we account only two normal modes (COM and stretch) for the selected segment (ions 3 and 4). In simulations we use a fixed gate duration of $200$ $\mu$s and the optical parameters taken from \cite{Leibfried2003, Matteo2020}:  two $313.2$ nm Raman lasers and the $30^{\circ}$ angle between the laser beams and the trap axis. The absolute magnitude of the Rabi frequency on each time interval does not exceed $2\pi \times 0.22$ MHz.

Second, we account for the phonon frequencies drift assuming that it linearly increases with time with rate $\gamma$ ranging from 100 Hz/min to 1 MHz/min:
\begin{equation}\label{Eq:drift_rate}
    \omega_m(t) = \omega_m + \gamma t.
\end{equation}
To include this drift, we divide the laser-pulse intervals on smaller sub-intervals characterized by its own constant normal frequency calculated according to Eq.\eqref{Eq:drift_rate}. We vary the number of sub-intervals to obtain the lower estimate of the gate fidelity. 
 
Finally, we calculate the fidelity of the gate using the following analytical formula \cite{Zhu2006, Figatt2018}:
\begin{eqnarray}\label{eq:MS_gate_fid}
\mathcal{F} = \frac{1}{8}\left [ 2+2(\Gamma_i + \Gamma_j)\cos{2\Delta\chi} + \Gamma_+ + \Gamma_- \right ],
\end{eqnarray}
where 
\begin{equation}
\begin{aligned}
&\Delta\chi = \pi/4 - \chi_{i,j}(t_g),\\
&\Gamma_{i,j} = \exp{-2\sum_m|\alpha_{i,j}^m(t_g)|^2\beta_m}, \\
&\Gamma_\pm = \exp{-2\sum_m|\alpha_i^m(t_g)\pm \alpha_j^m(t_g)|^2\beta_m},\\
&\beta_m = \coth{\dfrac{\hbar\omega_m}{2k_BT}} = \coth{\left [\dfrac{\omega_m}{2\omega_c}\ln{\left (1+\dfrac{1}{\bar{n}_c}\right )}\right ]}.
\end{aligned}
\end{equation}
Functions $\alpha_i^m(t_g)$ with $t_g$ as the gate time are respectful for the entanglement between ion $i$ and mode $m$, while the function $\chi_{i,j} (tg)$ accounts for the entanglement between ions $i$ and $j$ (see Appendix \ref{appendix:MSevolution} for more details). The factor  $\beta_m$ accounts for the initial effective normal mode temperature T and the initial average phonon occupation number of the ion pair for their COM mode $\bar{n}_c= (e^{\hbar\omega_c/k_BT}-1)^{-1}$ , $k_B$ is the Boltzmann constant. Eq. \eqref{eq:MS_gate_fid} was derived on the assumption that the initial internal state of the pair of ions is $\ket{\downarrow\downarrow}_{z}$.

\begin{figure}[t]
\includegraphics[width=3.375in]{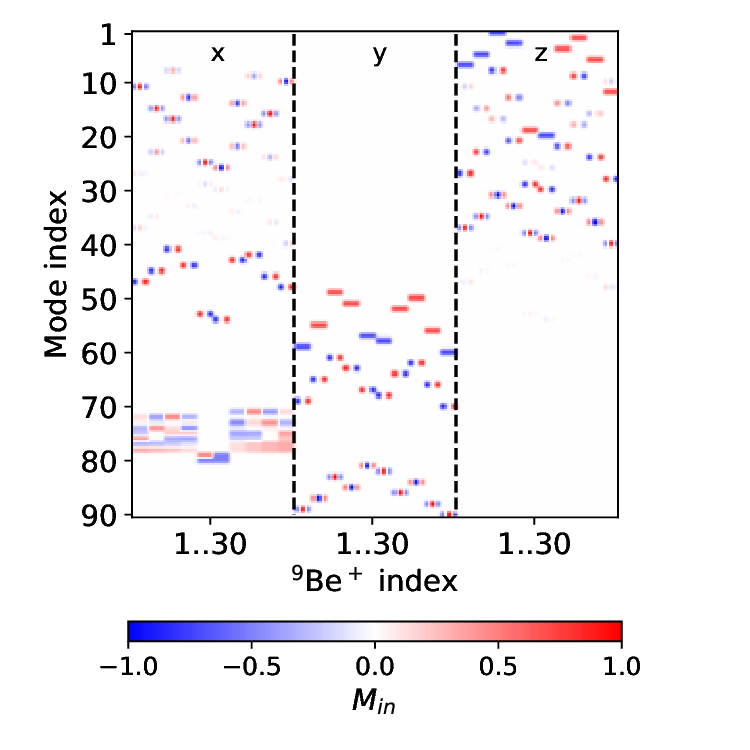}

\caption{\label{fig:Bew28ion3N} Interaction matrix for the trap with ten trapping sites, containing three Be ions in each site, similar to configuration \ref{fig:Bew28ion3axialcenpin}(b). The mode indexes are ordered according to their frequencies: the lower the frequency, the higher the index.}
\end{figure}

Fig. \ref{fig:RFreqsHeatmap}(b) shows results for Be ions, for different detunings $\mu$, and for different $\bar{n}_c$. Here we account for the entire phonon mode spectrum and the corresponding $\alpha_{i,m}(t_g)$-s.  $\mu$ ranges from $22.4$ to $22.45$ MHz with the step of $2$ Hz. Infidelity has a narrow maximum when $\mu$ is close to the normal mode frequencies or to half of their sum (see Fig. \ref{fig:RFreqsHeatmap}(b)). The maximum fidelities are achieved, when the lasers are detuned from the local infidelity maxima by at least 10 Hz. At $\gamma = $ 1 MHz/min the leading contribution to infidelity is the remaining ion-phonon entanglement, while for the drift rates below 150 kHz/min the leading contribution comes from the presence of the whole phonon-mode spectra. For the latter case, for the phonon mode occupation $\bar{n}$ of 20 the maximum infidelity reaches $5.7\cdot10^{-6}$. 

We also model parallel MS operations for Be qubits. For this we select two neighboring segments depicted in Fig. \ref{fig:Bew28axialladder}. The interaction between ions from two neighboring segments is three orders of magnitude weaker than between the ions within a segment, which leads to the independent gate operation. Indeed, we found that MS gate fidelity of two parallel gates on two adjacent segments is well approximated by multiplication of fidelities of two independent operations. Therefore, we estimate the fidelity of the parallel two-qubit gates for the entire chain to be greater than 99.95\%. This result can be further improved using the multi-loop gates and dynamical decoupling pulse sequences \cite{Milne2020, Hayes2012}.

\begin{figure}[b]
\begin{tabular}{cc}
\includegraphics[width=1.65in]{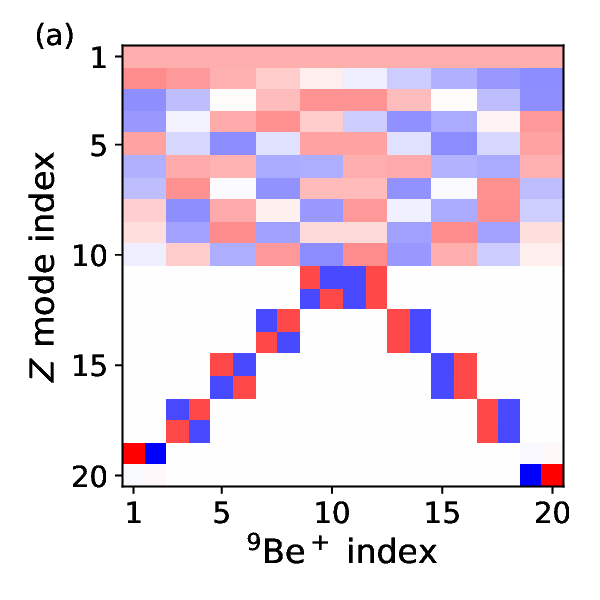}&
\includegraphics[width=1.65in]{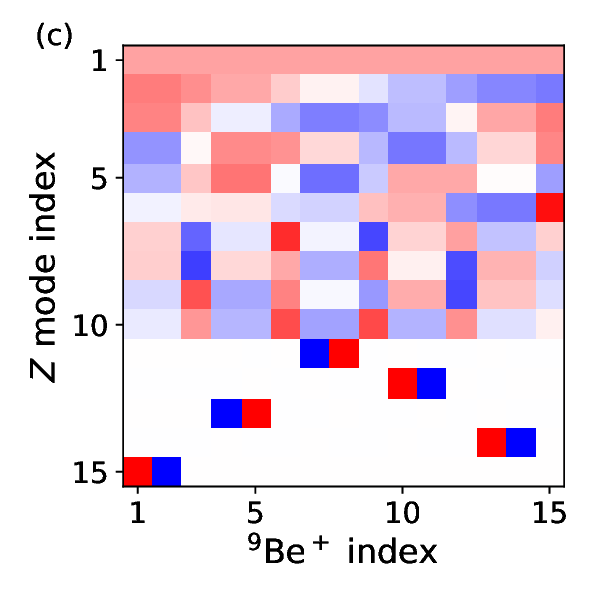}
\\
\includegraphics[width=1.65in]{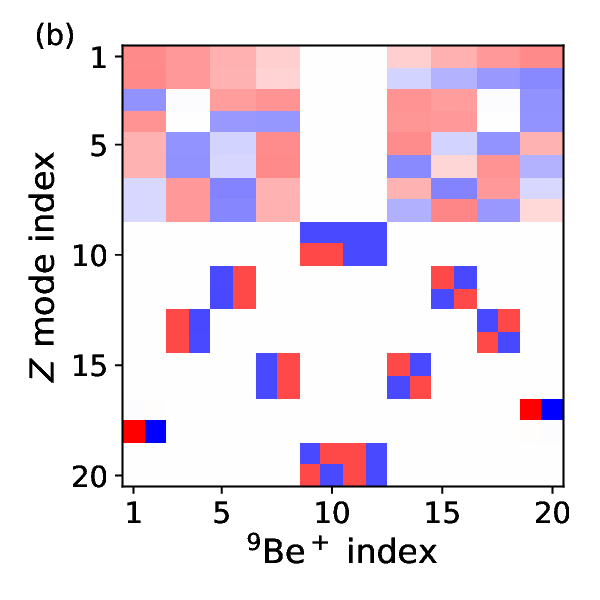}&
\includegraphics[width=1.65in]{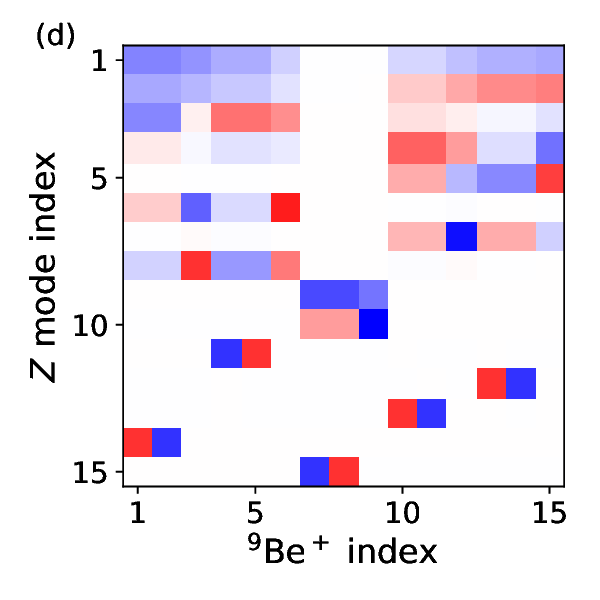}\\
\end{tabular}

\begin{tabular}{r}
\includegraphics[width=3in]{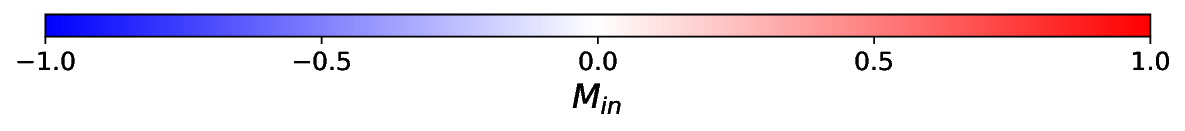}
\end{tabular}

\begin{ruledtabular}
\begin{tabular}{lccccc}
Mode index ($n$) & 1 & 9 & 11 & 19 \\
\colrule
(a) $f_{n}$ (MHz) & 61.4 & 61.4 & 56.7 & 56.3\\
(b) $f_{n}$ (MHz) & 61.4 & 58.3 & 56.7 & 53.6 \\
(c) $f_{n}$ (MHz) & 61.4 & 61.4 & 56.7 & --\\
(d) $f_{n}$ (MHz) & 61.4 & 61.4 & 56.7 & --\\
\end{tabular}
\end{ruledtabular}\\

\caption{\label{fig:Bew28ion2} Interaction matrices of 20 and 15  $^{9}$Be$^+$ ions sitting in the trap with 10 trapping sites shown in Fig. \ref{fig:Trapwithfield}. The colors correspond to the normalized strength of the ion-to-mode interaction calculated according to Eq. \eqref{eq:intermatrix}. Panels (a) and (c) demonstrate all-to-all coupling similar to Fig. \ref{fig:Caw28cenpin}(b) with the same set of dc voltages applied. Panels (b) and (d) demonstrate the interaction between the two centered traps similar to Fig. \ref{fig:Caw28cenpin}(d) with the same set of dc voltages applied. On the table the frequencies $f_n$ of the respective normal modes $n$ are presented. }
\end{figure}

\begin{figure*}
\begin{tabular}{ccc}
 \includegraphics[width=2.2in]{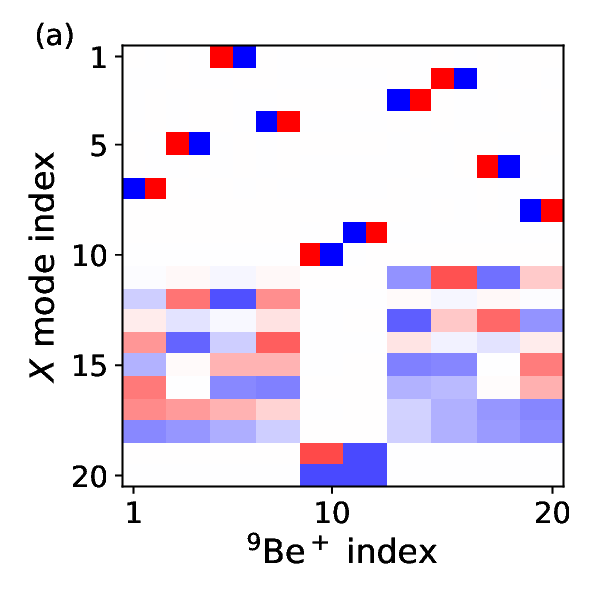}
&
\includegraphics[width=2.2in]{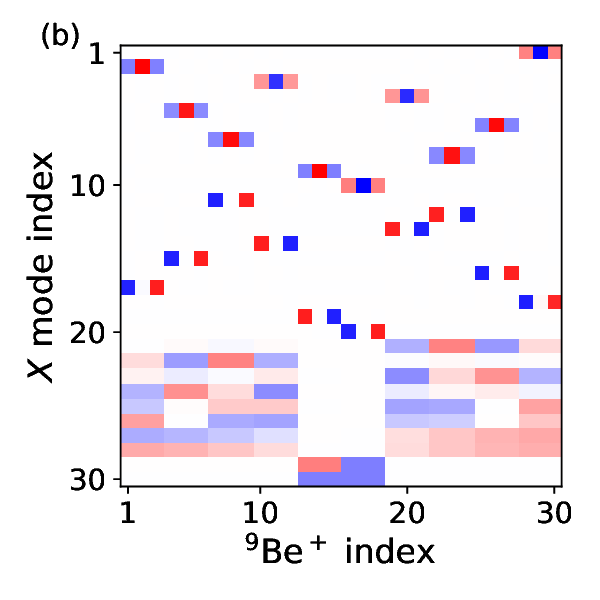}
& 
\includegraphics[width=2.2in]{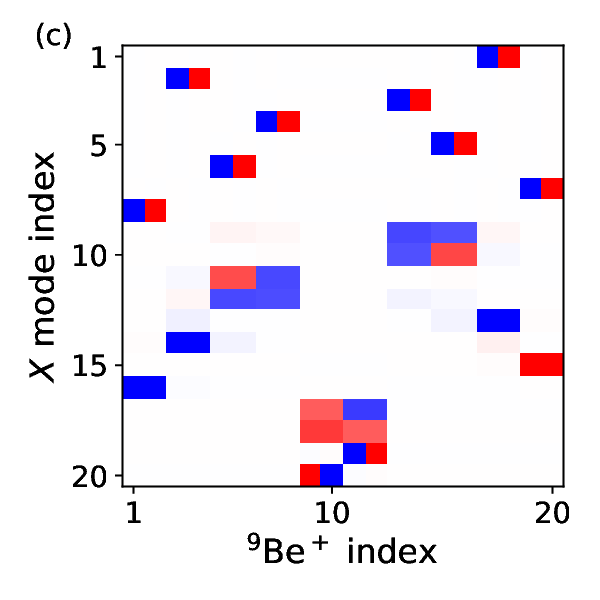}
\end{tabular}
\includegraphics[width=5in]{colorbar.eps}

\begin{ruledtabular}
\begin{tabular}{lccccccccc}
Mode index ($n$) & 1 & 7 & 10 & 12 & 16 & 18 & 20 & 28 & 30 \\
\colrule
(a) $f_n$ (MHz) & 40.8 & 40.6 & 38.6 & 22.9 & 22.9 & 22.9 & 21.7 & -- & --  \\
(b) $f_n$ (MHz) & 57.3 & 56.9 & 54.1 & 40.7 & 40.7 & 40.4 & 38.6 & 22.9 & 21.7  \\
(c) $f_n$ (MHz) & 40.6 & 40.6 & 22.5 & 22.5 & 22.3 & 21.2 & 38.4 & -- & --  \\
\end{tabular}
\end{ruledtabular}

\caption{\label{fig:Bew28ion3axialcenpin} 
Interaction matrices of 20 (Panel (a)) and 30 (Panel (b)) $^9$Be$^+$ ions confined in the trap with 10 trapping sites with two central sites united, calculated in a harmonic approximation. Panel (c) demonstrates an uncoupled interaction matrix of a configuration from panel (a) in the presence of octopole anharmonicity. For all panels the applied dc voltages are the same.}
\end{figure*}

\section{Coupling enhancement}
\label{coupling}

In this section, we investigate the capability to increase the coupling between the trapping sites using several ions per site and anharmonicity of the trapping potential. The positive effect of these two aspects on coupling strength has been demonstrated for two individual wells \cite{harlander2011trapped}. First, we examine the normal mode spectra for different numbers of ions in harmonic approximation and only then add anharmonic terms. 

\subsection{Harmonic approximation}

We choose the magnitude of the anisotropy parameter $\alpha = \omega_{x}^2/\omega^2_{y}=0.34$ to keep the ion crystal with up to three ions (Ca and Be) per trapping site linear along the $x$-axis (the critical value for 3 ions is $\alpha_{cr} = 0.38$ \cite{Yan2016}). Along  $z$ direction, the ion heights alternate with the magnitude $\delta z_{i}/z_{mean} \leq 0.0015$, which introduces a finite motional coupling between $x$- and $z$-axis, shown in Fig. \ref{fig:Bew28ion3N}, which complicates parallel gate implementation. 
This coupling can be characterized using the following relation:
\begin{equation}
    \frac{b_{i,n}^{secondary}}{b_{i,n}^{main}} \leq 0.2,
\end{equation}
where $b_{i,n}$ are the elements of the transformation matrix elements in Eq. \eqref{eq:normalmode} with captions ${main}$ and $secondary$ distinguishing the principal axes of motion depending on the magnitude of $b_{i,n}$ ( it is larger for the main axis). 
On the other hand, along each of the axes there are modes for which all three principal axes can be considered independently, for example, modes with indices 70-80 for x-axis. Therefore, further we present the part of the spectrum containing only these independent modes. 

Overall, the full crystal's motional spectrum is mediated by the ion motion within a trapping site due to the large difference between inter-ion distances within trapping sites and the nearest sites: $d_{inner}/d_{outer} = 1/28$. As a result, the inter-ion interaction within a single trapping site is $\sim10^3$ times stronger than between adjacent sites. For a trap with $N$ ions, with a mean number of ions per site $N_{site}$, we observe a separation of the mode spectrum for a chosen axis  into two parts. First part consists of $N/N_{site}$ modes (for example modes 1-10 in Fig. \ref{fig:Bew28ion2}(a)) and is similar to the case with a single ion per trapping site. The ions within a trapping site in these modes move in phase similar to the COM mode, but the phases of motion between different sites are not necessarily equal. We will refer to this part of the spectrum as individual-COM modes. The second part of the spectrum (modes 11-20) is similar to the spectrum in Fig. \ref{fig:Caw28dc03N} as if the trapping sites were uncoupled. These modes have frequency separation about 1 Hz, which is too small to perform parallel gates.
Therefore, the only modes suitable for parallel gates are the individual-COM modes. 

First, we select radial direction (along $z$ axis), unaffected by anharmonic terms, and examine how the number of Be ions in individual wells $n$ influences the phonon mode spectrum and coupling strength.  Fig. \ref{fig:Bew28ion2} shows interaction matrices for 20 and 15  $^{9}$Be$^+$ ions sitting in the trap with 10 trapping sites shown in Fig. \ref{fig:Trapwithfield}. Two and one ions per site in a 15-ion chain are alternated as follows: 2-1-2-1-2-... Panels (a) and (c) demonstrate all-to-all coupling similar to Fig. \ref{fig:Caw28cenpin}(b). Panels (b) and (d) demonstrate the interaction between the two centered traps similar to Fig. \ref{fig:Caw28cenpin}(d). Importantly, for the alternating ion number, the interaction between individual wells is still present, demonstrating the possibility to operate the trap with an unequal amount of ions in each trapping site.
  
\begin{table*}
\caption{\label{tab:table1} Coupling strength between Be ions in the central trapping sites (marked in bold) for the various populations of the ten trapping sites. For the results in the table, electrode voltages were kept the same for each of the axes in the simulations.}
\begin{ruledtabular}
\begin{tabular}{lcccccc}
\multicolumn{7}{c}{\textrm{Parameters of the simulations}}\\
\colrule
\colrule
\textrm{Population of the trapping sites}&
\multicolumn{2}{c}{1-1-1-1-\textbf{1-1}-1-1-1-1} &
\textrm{2-1-2-1-\textbf{2-1}-2-1-2-1}&
\multicolumn{2}{c}{\textrm{2-2-2-2-\textbf{2-2}-2-2-2-2}}&
\textrm{3-3-3-3-\textbf{3-3}-3-3-3-3}\\
\colrule
\textrm{Axis of Interaction}&
\textrm{$z$}&
\textrm{$x$}&
\textrm{$z$}&
\textrm{$z$}&
\textrm{$x$}&
\textrm{$x$}\\
\colrule
\textrm{Reference Fig.}&
\multicolumn{2}{c}{\textrm{Similar to Fig. \ref{fig:Caw28cenpin}(d)}}&
\textrm{Fig. \ref{fig:Bew28ion2}(d)}&
\textrm{Fig. \ref{fig:Bew28ion2}(b)}&
\textrm{Fig. \ref{fig:Bew28ion3axialcenpin}(a)}&
\textrm{Fig. \ref{fig:Bew28ion3axialcenpin}(b)}\\
\colrule
\textrm{Interacting mode indices ($n$)}&
\multicolumn{2}{c}{-}&
\textrm{9-10}&
\textrm{9-10}&
\textrm{19-20}&
\textrm{29-30}\\
\colrule
\colrule\\
\multicolumn{7}{c}{$\Delta f$ (Hz)}\\
\colrule
\colrule
Harmonic potential & 295  & 1528 &  493 & 587  &  3039 & 4589\\
Anharmonic potential & -  & -  &  - & -  & 3208 & 4998\\

\end{tabular}
\end{ruledtabular}
\end{table*}

Figs. \ref{fig:Bew28ion3axialcenpin}(a-b) demonstrate interaction matrices for 20 and 30 $^9$Be$^+$ ions in axial direction in a harmonic approximation. Each of the ten trapping sites holds an equal number of ions: two (Fig. \ref{fig:Bew28ion3axialcenpin}(a)) or three (Fig. \ref{fig:Bew28ion3axialcenpin}(b)). The two central traps are united in terms of their spectrum (modes 19-20 and 29-30, respectively). 

The coupling strengths between two united trapping sites for different ion configurations in harmonic approximation are listed in Table \ref{tab:table1}. As expected from the previous work \cite{harlander2011trapped}, in  harmonic approximation, coupling strength grows linearly with the number of ions per trapping site. Further we investigate the anharmonic effects on coupling strength. 

\subsection{Anharmonic potential}

To account strong anharmonic effects in axial direction we expand the trapping site potential near its minimum along $x$ axis up to the fourth order \cite{Home_2011}:
\begin{equation}
\begin{aligned}
U_{anh}(\delta x) = \sum^{\infty}_{n=2}\kappa_n\delta x^n\approx \kappa_2\delta x^2\left[ 1 + \left(\frac{\delta x}{\lambda_3} \right)+\left(\frac{\delta x}{\lambda_4} \right)^2 \right].
\end{aligned}
\end{equation}
 Here $l = (Ze/8\pi\epsilon_0\kappa_2)^{1/3}$ is a characteristic length of the potential, and $\lambda_n = (\kappa_n/\kappa_2)^{1/(2-n)}$ is anharmonicity scale length. For both 
 Be and Ca ions $l=0.74$ {\textmu}m, $\lambda_3 = 2.4\times 10^3$ {\textmu}m and $\lambda_4 = 7.2$ {\textmu}m. Thus, the fourth order term dominates the third order one which we account in the Jacobian $A_{mn}$ as follows: 
\begin{equation}
\begin{aligned}
A_{mn} = 
\begin{cases}
    \dfrac{\kappa_2^n}{\kappa_2^o}+6\theta_n \left(\delta u_n\right)^2 + 2\displaystyle\sum^{N}_{\substack{p=1 \\ p\neq n}}\frac{1}{|u_n - u_p|^2} \text{ if } n\neq m,\\ 
    \dfrac{-2}{|u_n - u_m|^2} \text{ if } n = m.
\end{cases}
\end{aligned}
\end{equation}
Here $n$ is the ion index and we introduced the dimensionless coordinates $u_n = x_n/l$, $\theta_n = l^2\kappa_4^n\kappa_2^o / \left( \kappa_2^n\right)^2$ characterizes octopole anharmonicity of the trap and $\kappa_2^o$ is a curvature of the potential in a trapping site with the highest frequency. The equilibrium positions (potential minima) are determined through simulation of the ion motion in the anharmonic time-dependent potential. The shifted normal mode frequencies are calculated from the eigenvalues of the Jacobian. Then the coupling strengths between two coupled wells are obtained out of the differences between them. Table \ref{tab:table1} summarizes the coupling strengths between ions in two united trapping sites. The impact of the anharmonic terms increases with the number of ions per site. 

Fig. \ref{fig:Bew28ion3axialcenpin}(c) shows the interaction matrix for Be ions in the anharmonic trapping potential. The rest parameters of the simulations were the same as in Fig. \ref{fig:Bew28ion3axialcenpin}(a). The matrices look very different due to the variation of the anharmonicity with the trapping site index: $\theta$ ranges  $\sim[-0.014, -0.021]$. 
This complicates the formation of the ladder-type structure in Fig. \ref{fig:Bew28axialladder}(b) required for parallel gate implementation. The effect of anharmonic potential becomes visible when the coupling strength between trapping sites is well below normal mode frequency shifts.
Overall, it can be compensated by the proper adjustment of the dc voltages. Details on this procedure are beyond the scope of this work.

\section{Conclusion and outlook}
In conclusion, we described and characterized a new surface trap design allowing to connect or disconnect different sites of the trap and allowing implementation of parallel gate operations. In particular, we demonstrated that by changing the dc voltages on the trapping electrodes we can unite the chosen ions into segments with unique phonon mode frequencies. For this we modeled the trapping potential and the corresponding phonon mode spectra for Ca and Be ions. We found optimum configuration for parallel gates and verified the performance of the parallel MS-gate operation for it. The infidelity of the gate was calculated under the drift of the normal mode frequencies expected in the experiment and for different finite populations of the normal modes. We show that for experimental drift rates of a few kHz per several minutes the infidelity from the normal frequency drift does not exceed $5\cdot 10^{-6}$ for a wide range of laser detunings. For higher drift rates the ion-to-mode entanglement starts to play an important role. We calculated the fidelity of the gate of 99.97\% for a drift rate $\gamma = 1$ MHz/min and initial average COM mode occupation number $\bar{n}_c = 20$. 

We discussed the scaling of the proposed design and verified it for 50 trapping sites. The proposed trap design was also tested for two and three ions per trapping site. In general, the coupling strength increases linearly with the number of ions and, therefore, each segment can comprise any reasonable number of ions. Anharmonic terms of the surface trap potential decouple the non-symmetrical individual wells, which limits the possible number of ions per trapping site. 

The proposed design opens the possibility to construct the nearest-neighbor interactions without all-to-all connectivity, which are naturally absent in the conventional Paul traps. Such couplings are advantageous for digital quantum computations, since they allow parallel gate implementation and therefore speedup. For example, such couplings might be appealing for variational algorithms with the ansatz structures like HEA (hardware efficient ansatz). Importantly, the proposed trap architecture does not exclude the possibility to entangle distant ions or several ions at once (as illustrated on Fig. \ref{fig:Caw28cenpin}) additionally offering the possibility to tune the strength of those couplings. The latter has the potential to broaden the potential area of quantum simulations involving trapped ion strings. All in all, we expect the proposed design to be competitive with the present scaling methods in respect of achieving large-scale quantum computer.

\section*{Acknowledgements}
We would like to thank Andrei Chuchalin, Mikhail Popov, Alexey Zaytsev and Stepan Kartsev for insightful discussions.
This work was supported by Rosatom in the framework of the Roadmap for Quantum computing (Contract No. 868-1.3-15/15-2021 dated October 5).

\appendix
\section{Ion dynamics simulation}
\label{appendix:simulation}

We have developed a Python package \emph{Sion} \cite{Podlesnyy_Sion_Python_package} capable of accounting sophisticated surface trap electrode structure. It is based on the pyLIon package \cite{BENTINE2020} for ion dynamics simulation in 3D Paul traps and allows obtaining ion trajectories and calculating normal mode spectra. 
pyLIon in its turn utilizes the LAMMPS software employing the molecular dynamics approach to describe the motions of large systems with many particles. Numerical integration of Newton’s classical equations of motion of each confined atom is performed through the Verlet algorithm \cite{doi:10.1063/1.442716}. 
Laser cooling of ions and background gas collisions are modeled using the Langevin equation:
\begin{equation}
    m_i\frac{d^2\boldsymbol{x}}{dt^2} = F_i(\boldsymbol{x}) - \gamma_i\dot{\boldsymbol{x}} + f_i(t),
\end{equation}
where $m_i$ is the ion mass, $F_i$ is the force of the trapping potential acting on ion $i$ experiencing the cooling force with strength $\gamma_i$ and stochastic force $f_i$. The following total potential is used to determine the force $F_i$ for $N$ ions confined in the trap:
\begin{equation}
\begin{aligned}
&U_{pot}(x,y,z,t)=\\
&\sum^{N_{DC}}_{i=1}\phi^i_{DC}(x,y,z) 
    +\sum^{N_{RF}}_{i=1}\phi^i_{RF}(x,y,z)\cos{(\Omega_{RF} t)}\\
   & + \sum^{N}_{n = 1}\sum^{N}_{m > n} \frac{Z^2e^2}{4\pi \epsilon_0}\frac{1}{ \vert\vec{r}_{n} - \vec{r}_{m} \vert^2}, 
\end{aligned}
\end{equation}
 with $N_{DC}$, $N_{RF}$ -- the number of dc and RF electrodes, correspondingly. It includes both the force originating from electric potential and from the mutual Coulomb interaction between the ions (last term). Sion defines electric force acting on the ion from the electrode chip. According to the concept of superposition, the total potential above the electrodes can be obtained by adding the potential generated by each electrode.

To model electric field above the electrode surface and for the subsequent calculation of the equilibrium ion positions we used analytical form of the electric field produced by rectangular surface electrode under unit voltage from \cite{House2008}:

\begin{equation}
\begin{aligned}
    &\phi(x,y,z) = \\
    &\frac{V}{2\pi}
    \bigg[ 
    \tan^{-1}\left( \frac{(x_b-x)(y_b-y)}{z\sqrt{z^2+(x_b-x)^2+(y_b-y)^2}} \right) \\ 
    &-\tan^{-1}\left( \frac{(x_a-x)(y_b-y)}{z\sqrt{z^2+(x_a-x)^2+(y_b-y)^2}} \right) \\ 
    &-\tan^{-1}\left( \frac{(x_b-x)(y_a-y)}{z\sqrt{z^2+(x_b-x)^2+(y_a-y)^2}} \right) \\ 
    &+\tan^{-1}\left( \frac{(x_a-x)(y_a-y)}{z\sqrt{z^2+(x_a-x)^2+(y_a-y)^2}} \right)
    \bigg].
    \end{aligned}
\end{equation}
Here $V$ is the applied voltage and $(x_a,y_a, 0)$ and $(x_b, y_b, 0)$ are the coordinates of two opposing vertices of the planar electrode. 

Simulation starts with the initialization of the ion coordinates and velocities. In our case we assume ions at the geometrical center of each trapping site and set their initial velocity to zero.
The equilibrium positions are then obtained by monitoring ion dynamics and coordinates, setting the stochastic force $f_i(t) = 0$.
The benefit of such a method is a possibility to account for the exact form of the potential and to determine the stability of the system.

\begin{figure}[t]
\includegraphics[width=3.375in]{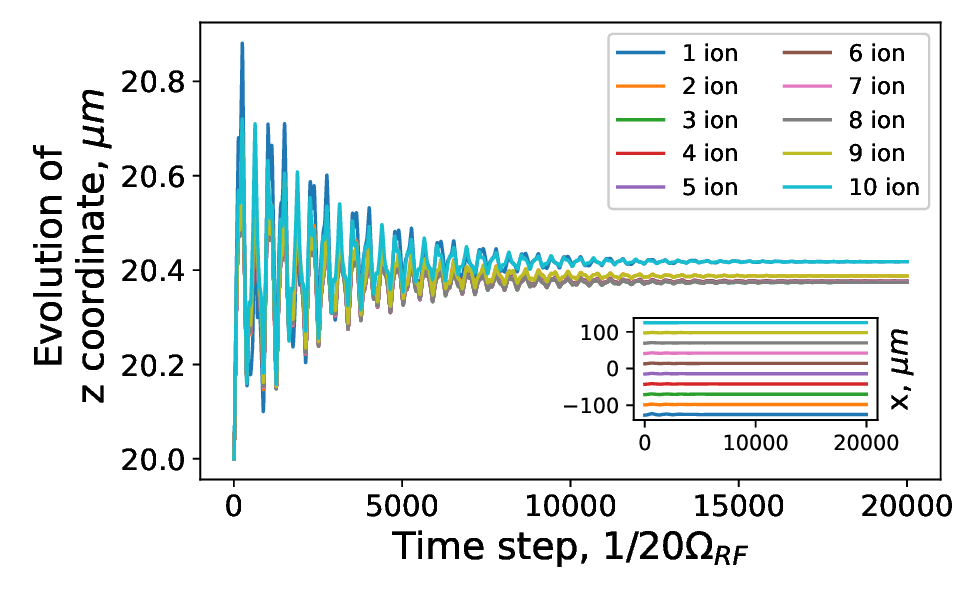}
\caption{\label{fig:Caw28dc0evolz} Simulated evolution of z and x (in the inset) coordinates of 10 trapped ions in the presence of cooling simulated for the trap in Fig. \ref{fig:Trapwithfield} and grounded dc electrodes. Time step is $1/20\Omega_{RF}$. The variation between the heights of different ions at equilibrium is about $1.5\times10^{-3}z_{\text{mean}}$. }
\end{figure}

The result of the simulation is presented as a trajectory of each ion. Observing the trajectory, we are able to determine, if the ion crystal is stable, and identify the moment, when the crystal is cooled to 0 K. The time-evolution of $z$- an $x$-coordinates of ten trapped-ions for the grounded DC-electrodes (configuration from Fig. \ref{fig:Caw28dc03N}) is demonstrated in Fig. \ref{fig:Caw28dc0evolz}. The simulation time is given in time steps since they are relative to $\Omega_{RF}$.

Knowing the equilibrium ion positions, masses and secular frequencies (see Eq.\eqref{eq:hessian}) we calculate the normal modes. For this, we introduce several notations: the dimensionless coordinates $(u_i,v_i,w_i) = (x_i/l,y_i/l, z_i/l),$ $l^3 = e^2/(4\pi\epsilon_0m_o\omega_{zo}^2)$, where $m_o$ is the mass of the ion with index \emph{1}, and $\omega_o$ its secular frequency; a $3N\times 3N$ mass matrix $M$ defined as $M = \text{Diag} [m_1..m_N,m_1..m_N,m_1..m_N]$, where $m_i$ is a mass of the ion with index \emph{i}; $\rho = m_i\omega_{ki}^2/m_o\omega_{ko}^2$, where $k = x, y, z$ - denotes the principal axis; a dimensionless distance between ions \emph{i,j} $d_{ij} = \sqrt{(u_i-u_j)^2+(v_i-v_j)^2+(w_i-w_j)^2}$.
For a given equilibrium ion positions  Sion computes the hessian matrix in dimensionless coordinates $A_{mn}$ through the following equations:

\begin{align*}
    & A_{ii}  = \rho_{xi} + \sum^N_{p=1}\left( \frac{-1}{d_{ip}^3} + \frac{3(u_i-u_p)^2}{d_{ip}^5}\right), \\
    & A_{(N+i)(N+i)}  = \rho_{yi} + \sum^N_{p=1}\left( \frac{-1}{d_{ip}^3} + \frac{3(v_i-v_p)^2}{d_{ip}^5}\right), \\
    & A_{(2N+i)(2N+i)}  = \rho_{zi} + \sum^N_{p=1}\left( \frac{-1}{d_{ip}^3} + \frac{3(w_i-w_p)^2}{d_{ip}^5}\right), \\
    & A_{ij} = \frac{1}{d_{ij}^3} - \frac{3(u_i-u_j)^2}{d_{ij}^5}, \\
    & A_{(N+i)(N+j)} = \frac{1}{d_{ij}^3} - \frac{3(v_i-v_j)^2}{d_{ij}^5},\\
    & A_{(2N+i)(2N+j)} = \frac{1}{d_{ij}^3} - \frac{3(w_i-w_j)^2}{d_{ij}^5},\\
    & A_{i(N+j)} = \frac{3(u_i-u_j)(v_j-v_i)}{d_{ij}^5},\\
    & A_{i(2N+j)} = \frac{3(u_i-u_j)(w_j-w_i)}{d_{ij}^5},\\
    & A_{(2N+i)(N+j)} = \frac{3(w_i-w_j)(v_j-v_i)}{d_{ij}^5},\\
    & A_{i(N+i)} = \sum^N_{p=1}\frac{3(u_i-u_p)(v_i-v_p)}{d_{ip}^5},\\
    & A_{i(2N+i)} = \sum^N_{p=1}\frac{3(u_i-u_p)(w_i-w_p)}{d_{ip}^5},\\
    & A_{(N+i)(2N+i)} = \sum^N_{p=1}\frac{3(w_i-w_p)(v_i-v_p)}{d_{ip}^5},
\end{align*}
, where $i,j = 1..N$. The $3N\times 3N$ matrix $A_{mn}$ is then fully defined by these equations, since as a hessian it is symmetric.
The normal mode vectors $b_{mi}$ are then eigenvectors and squared normal frequencies $f^2_m$ are eigenvalues of matrix $D$:
\begin{equation}
    D = m_o\omega_o^2M^{-1/2}A_{mn}M^{-1/2}.
\end{equation}
\\
\section{DC voltage set optimization}
\label{appendix:DCoptimization}

A strict condition on the secular frequency matching between different trapping sites is necessary for their coupling (see section \ref{sec:spectral}) and requires a numerical optimization procedure for the voltage set. We define an optimization problem as a task to find a minimum of a loss function:
\begin{equation}
    L(V^{DC}_{set}) = \sum^{N_{DC}}_{i=1}\left( \omega^{actual}_i - \omega^{desired}_i \right)^2.
\end{equation}
Here, the loss function defines a quadratic norm of difference between the actual secular frequency $\omega^{actual}_i$ on each trapping site $i$, and the desired frequency $\omega^{desired}_i$. The actual frequencies are calculated for the given voltage set on the central dc electrodes $V^{DC}_{set}$. The optimization parameter space is $N_{DC}$-dimensional, for the number of central dc electrodes $N_{DC}$. The frequencies are normalized to improve the algorithm performance. 

The numerical procedure, considering the secular frequencies in each trapping site for a given voltage set is necessary.
The operating voltages on each dc electrode range from 0 to 6 V. This restricts the amplitude of the variation of secular frequencies. Namely, suppose $\delta\omega = \omega(0) - \omega(6)$, where $\omega(V)$ denotes a secular frequency in a trapping site, with corresponding dc voltage $V$, and with all the other dc electrodes being grounded.
We estimate the following variations: $(\delta\omega_x/\omega_x(0),\delta\omega_y/\omega_y(0),\delta\omega_z/\omega_z(0))$ = (12.3, 9.4, 19.2)\%, for the Be ion and (11.6, 8.9, 18.1)\% for the Ca ion.
The variation of dc voltage on a trapping site influences the secular frequencies in all trapping sites. In particular, when the voltage of the dc electrode in the trapping site $i$ is varied from 0 V to 6 V, the secular frequency in the neighboring trapping sites $i\pm1$ shifts by $(\delta\omega_x/\omega_x(0),\delta\omega_y/\omega_y(0),\delta\omega_z/\omega_z(0))$ equal to (2.3, 0.2, 0.4)\% for Be and to (2.4, 0.2, 0.4)\% for Ca. The effect is stronger for the axial direction, which makes the optimization of voltage sets much more demanding.

We used the ADAM algorithm for optimization \cite{Kingma2014Adam}. The method for calculating gradients in each iteration varies for different use-cases. To obtain voltage configurations for Figs. \ref{fig:Caw28cenpin}, \ref{fig:Bew28axialladder} the exact gradient was calculated at each iteration.

The calculation of the secular frequency implies the additional search of the pseudopotential minimum of the trapping sites, which increases the time of each iteration of gradient descent. So for more complicated optimization problems, the number of calls to loss function must be reduced. This was achieved by using stochastic calculation of gradients for ADAM algorithm. The voltage configuration for the trap with 50 trapping sites in Fig. \ref{fig:Bew2850ladder} was calculated by this method. We observed that the stochastic choice of five dc voltages at each iteration to calculate the loss function provides the same convergence for 10-50 trapping site number. The execution time for 50 trapping sites is $\sim2$ hours on a conventional PC.

The voltage configuration in Fig. \ref{fig:heisenberg} was achieved by optimizing simultaneously the central and side dc electrode voltage set $V^{DC}_{set, whole}$. We choose the loss function as a quadratic norm of difference between actual frequencies and the desired set in all three oscillation axes:
\begin{equation}
    L(V^{DC}_{set, whole}) = \sum^{N_{DC, whole}}_{i=1}\sum^{3}_{k=1} \left( \omega^{actual}_{i,k} - \omega^{desired}_{i,k} \right)^2,
\end{equation}
where $k=x,y,z$ denotes the principle axis, $N_{DC, whole}$ -- the number of all electrodes -- central and side.
\\

\begin{table*}
\caption{\label{tab:asymmtric_coupling} The table demonstrates coupling strengths $\Omega_{I}$ obtained using Eq. \eqref{eq:interaction_fr} and theoretical coupling strengths $\Omega_{I}^{theor}$ calculated using Eq. \eqref{eq:interaction} for different normal mode configurations presented in this paper. The configurations are supported by reference figures of the respective interaction matrices. The trapping site indexes selected for the calculation are in the first column. The secular frequencies of those trapping sites match. $d_{ion}$ refer to the distance in {\textmu}m between the selected ions respectively.
}
\begin{ruledtabular}
\begin{tabular}{ccccccccccc}
Ion &
Site indexes & Reference Fig. & $d_{ion} $ ($\mu $m) & $\Omega_I/2\pi$ (Hz) & $\Omega^{theor}_{I}/2\pi$ (Hz) \\
\colrule \\
$^9$Be$^+$  & 5-6     &   Fig.\ref{fig:symmasimm}(d)             & 28                      & 1534                & 1534                          \\
&3-8     &     Fig.\ref{fig:symmasimm}(e)           & 140                     & 13.36                  & 13.22                            \\
&1-10    &     Fig.\ref{fig:symmasimm}(f)              & 252                     & 2.32                  & 2.27                            \\
&1-2     &     Fig.\ref{fig:Bew28axialladder}(b)           & 28                      & 1860                & 1692                          \\
&3-4     &        Fig.\ref{fig:Bew28axialladder}(b)        & 28                      & 1622                & 1656                          \\
&5-6     &       Fig.\ref{fig:Bew28axialladder}(b)         & 28                      & 1654                & 1622                          \\
&7-8     &       Fig.\ref{fig:Bew28axialladder}(b)         & 28                      & 1649                & 1589                          \\
&9-10    &      Fig.\ref{fig:Bew28axialladder}(b)          & 28                      & 1763                & 1557                          \\
\colrule \\
$^{40}$Ca$^+$ & 5-6     &        Fig.\ref{fig:Caw28cenpin}(d)        & 28                      & 141.83              & 148.55                   \\
&4-7     &       Fig.\ref{fig:symmasimm}(a)            & 84                      & 5.50                & 5.50                  \\
&3-8     &        Fig.\ref{fig:symmasimm}(b)           & 140                     & 1.19                & 1.19                   \\
&2-9     &         Fig.\ref{fig:symmasimm}(c)       & 196                     & 0.43                & 0.43                  \\
&1-10    &     Fig.\ref{fig:Caw28cenpin}(f)           & 252                     & 0.21                & 0.20               
\end{tabular}
\end{ruledtabular}
\end{table*}

\begin{figure*}[t]
\begin{tabular}{ccc}
\includegraphics[width=1.99in]{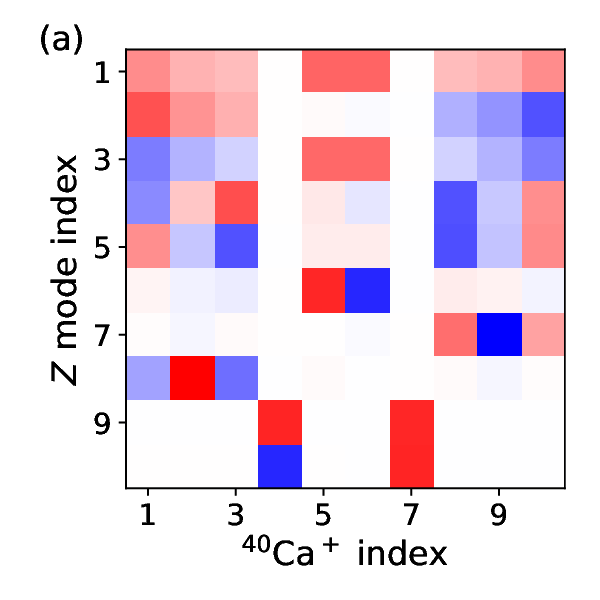}&
\includegraphics[width=1.99in]{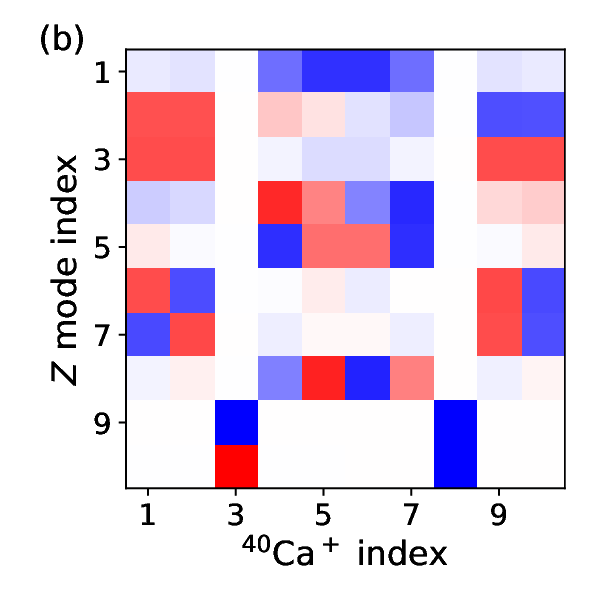}
& \includegraphics[width=1.99in]{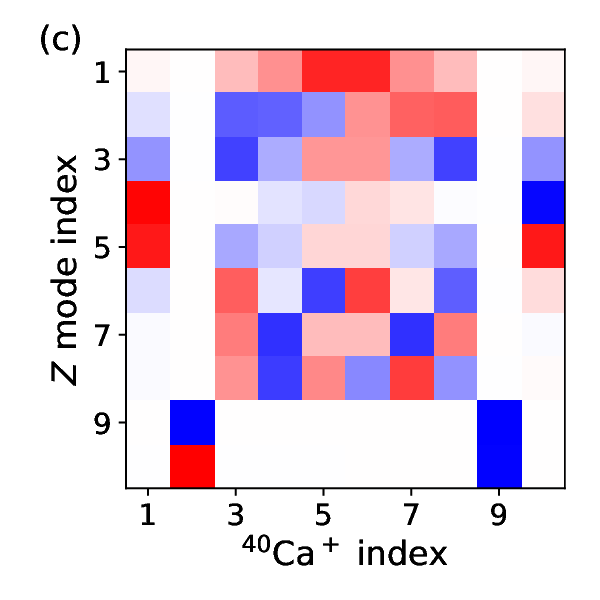}
\\
\includegraphics[width=1.99in]{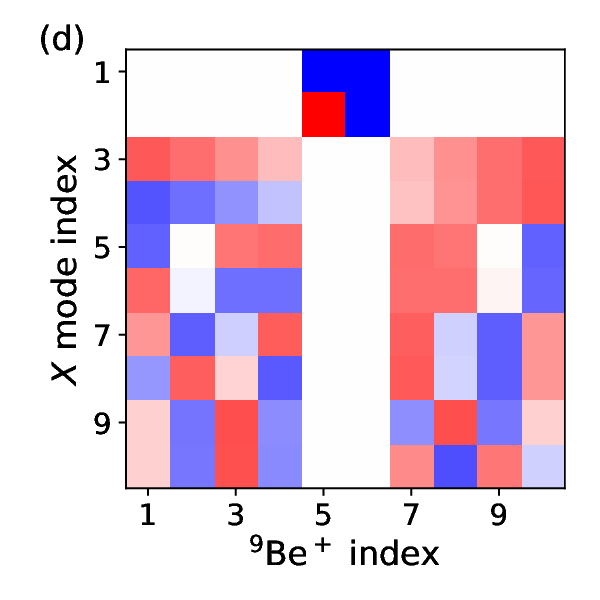}&
\includegraphics[width=1.99in]{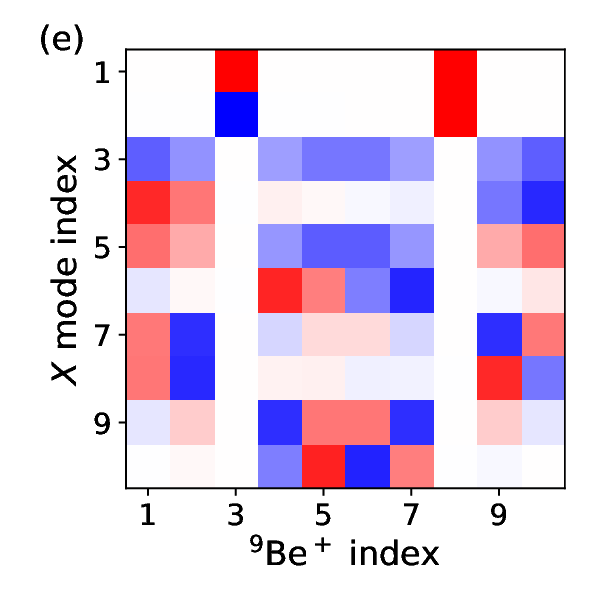}
& \includegraphics[width=1.99in]{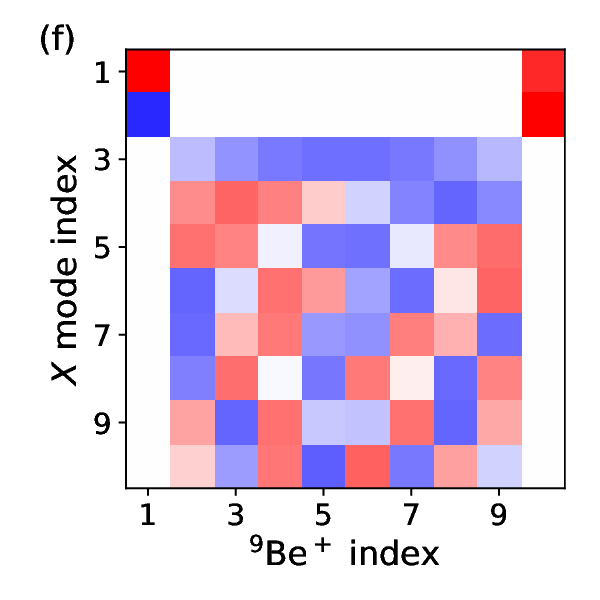}
\end{tabular}

\begin{tabular}{r}
\includegraphics[width=5in]{colorbar.eps}
\end{tabular}

\caption{\label{fig:symmasimm}  Normal mode interaction matrices. The first row demonstrates configurations for $Z$-mode spectral segments of Ca ions with indices (a) 4-7, (b) 3-8, (c) 2-9. The second row consists of configurations of $X$-mode spectral segments of Be ions with indices (a) 5-6, (b) 3-8, (c) 1-10.}
\end{figure*}

\section{Evolution operator for M$\o$lmer-S$\o$rensen gate}
\label{appendix:MSevolution}
We model the evolution of the state of the pair of qubits under the following operator:
\begin{equation}
\label{eq:Evolution operator general}
\begin{aligned}
{\hat U}(t) = exp\bigg [\sum_{i,m}&(\alpha_{i,m}(t)\hat a_m^\dagger -\alpha_{i,m}^*(t)\hat a_m)\hat\sigma_x^{(i)}\bigg ]\\
            &\times exp\bigg [i\sum_{i,j}\chi_{i,j}(t)\hat\sigma_x^{(i)}\hat\sigma_x^{(j)}\bigg ],\\
\end{aligned}
\end{equation}
where $\hat\sigma_x^{(i)}$ is the Pauli-$X$ operator acting on an ion $i$ and $\hat a_m^\dagger$/$\hat a_m$ are creation/annihilation operators of a phonon mode $m$. The functions $\alpha_{i,m}(t)$ are responsible for ion-phonon entanglement, while the phases $\chi_{i,j}$ account for entanglement between a pair of qubits $i,j$. 
The operator in Eq. \eqref{eq:Evolution operator general} was derived under the rotating wave approximation and in the Lamb-Dicke and resolved-sideband limits \cite{Choi2014, Zhu2006, Sosnova2021, Zhu2006EPL}. 

For the ideal gate implementation the following equations should be satisfied for Rabi frequencies on intervals:

\begin{eqnarray}
\label{eq: alpha and chi}
\alpha_{i,m}(t_g) = i\sum_{s=1}^{2N+1}\Omega_{s}\eta_{i,m}\int_{(s-1)t_p}^{st_p}sin(\mu t)e^{i\omega_{m}t}dt=0,
\end{eqnarray}
\begin{eqnarray}
\chi_{i,j}(t_g)&=&\sum_{r,s=1}^{2N+1}\Omega_{r}\Omega_{s}\Gamma_{rs,ij}=\pi/4,
\end{eqnarray}
where
\begin{eqnarray}
\begin{aligned}
\Gamma_{rs, ij}=&2\sum_{m=1}^N\int_{(r-1)t_p}^{rt_p}dt_2\int_{(s-1)t_p}^{min(st_p, t_2)}  dt_1\eta_{i,m}\eta_{j,m}\\
&\times
sin(\mu t_2)sin(\mu t_1)sin[\omega_{m}(t_2-t_1)].
\end{aligned}
\end{eqnarray}
Above $\Omega_s$ is the Rabi frequency on an interval $s$, $t_p = t_g/(2N+1)$ is the interval time, $\omega_m$ is the frequency of mode $m$, $\eta_{i,m}\propto b_{im}$ is the Lamb-Dicke parameter of ion $i$ in the mode $m$, and $min(a,b)$ depicts the minimum of two numbers $a$ and $b$. The rest parameters are the detuning of the frequency of addressing laser fields (Raman or optical) from the carrier transition $\mu$ which is close to the normal mode frequencies \cite{Choi2014, Zhu2006EPL}, and the gate duration $t_g$.
\\
\section{Coupling Hamiltonian}
\label{appendix:Coupling Hamiltonian}
We consider two singly charged ions of masses $m$ cooled to the motional ground state in individual harmonic traps separated by a distance $d$ with frequencies $\omega_{a}$, $\omega_{b}$ along the $x$-axis. The potential energy of ions expanded up to the second order on deviations of ions from equilibrium positions $x_a$, $x_b$ \cite{brown2011coupled} and with neglected constant energy terms can be expressed as follows:
\begin{equation}
\label{eq:potential}
\begin{aligned}
&U(x_a, x_b)=
\frac{m\omega_a^2x_a^2}{2} + \frac{m\omega_b^2x_b^2}{2} - \frac{e^2}{4\pi \epsilon_0 d} \frac{2 x_a x_b}{d^2}, \\
\end{aligned}
\end{equation}
The last term comes from the Coulomb repulsion and represents coupling between the ions’ motions. After quantisation  $x_{a} = \sqrt{\dfrac{\hbar}{2m\omega_{a}}} (a+a^\dagger)$, $x_{b} = \sqrt{\dfrac{\hbar}{2m\omega_{b}}} (b+b^\dagger)$ it becomes:
\begin{equation}
\begin{aligned}
&\frac{e^2}{4\pi \epsilon_0 d}\frac{x_a x_b}{d^2} = \hbar \Omega_{I} (a+a^\dagger)(b+b^\dagger),\\
&\Omega_{I} = \frac{e^2}{4\pi\epsilon_0 d} \cdot \frac{2}{m\omega_a\omega_bd^2} \sqrt{\omega_a\omega_b}.\\
\end{aligned}
\end{equation}
where $a/a^\dagger$ ($b/b^\dagger$) are annihilation/creation operators of ion $a$($b$) and $\Omega_{I}$ is the coupling strength. $\Omega_{I}$ describes the ratio between the Coulomb energy of ion interaction and the geometric mean of motional oscillator energies of two ions with displacement $d$. The full motional Hamiltonian of the ions in rotating wave approximation (RWA) with neglected constant terms then becomes \cite{wilson2014tunable}:
\begin{equation}\label{eq:quantised_ham}
\begin{aligned}
&H = \hbar\omega_a a^\dagger a + \hbar\omega_b b^\dagger b + 2\hbar \Omega_{I} (ab^\dagger+a^\dagger b).\\
\end{aligned}
\end{equation}

The normal vectors and frequencies for such a Hamiltonian take a form:
\begin{equation}\label{eq:quantised_ham}
\begin{aligned}
& \omega_{br/COM} = \bar{\omega} \pm \sqrt{(\Delta^{a,b})^2 + \Omega_{I}^2},\\
& b_{br/COM} = (\sin(\theta_{br/COM}),  \cos(\theta_{br/COM})),\\
\end{aligned}
\end{equation}
where 
\begin{equation}
  \label{eq:theta}
\theta_{br/COM} = \arctan{\left [ \frac{\Delta^{a,b} \mp \sqrt{(\Delta^{a,b})^2+\Omega_{I}^2}}{\Omega_{I}} \right ]},\\
\end{equation}
and $\bar{\omega} = \omega_a+\omega_b$, $\Delta^{a,b} = \omega_a - \omega_b$. 
From Eqs.~\eqref{eq:quantised_ham}, \eqref{eq:theta}, one can see that for large separation between secular frequencies ($\Delta^{a,b} \gg \Omega_I$) the motion
of two ions becomes decoupled. This makes it impossible to entangle the ions. In contrast, for $\Delta^{a,b} = 0$, 
the splitting between the normal mode frequencies equals $\Omega_{I}$ and the normal vectors become $b_{br/COM} = (\sqrt{2}/2, \pm\sqrt{2}/2)$ regardless 
of the distance between the traps. 
Thus, the motional coupling $\Omega_{I}$ can be used to characterize the possibility to entangle ions in separate wells, 
and in the case of equal secular frequencies of the wells it can be quantified as the difference between breathing and COM mode frequencies of the ion motion.

For a multi-trapping zone architecture, Eq. \eqref{eq:interaction_fr} is not necessarily valid. We verify numerically that for normal mode configurations studied in this paper, it works pretty well, as shown in Table \ref{tab:asymmtric_coupling}. We emphasize that the relation in Eq. \eqref{eq:interaction_fr} was used to find the optimum electrode configuration. The feasibility to perform MS gate in the proposed trap was verified by computing theoretical fidelities as described in the main text. 

\bibliography{citation}

\end{document}